\documentclass[
journal=cmatex,
layout=twocolumn,
manuscript=article,super=false]{achemso}

\usepackage[version=4]{mhchem} % Formula subscripts using \ce{}
\usepackage[dvipsnames]{xcolor}
\usepackage{siunitx}
\usepackage{subfig}
\usepackage{amsmath}
\usepackage{caption}
\usepackage{xfrac}
\usepackage[export]{adjustbox}

\usepackage{hyperref}

\hypersetup{
    colorlinks=true,
    linkcolor=red,
    filecolor=magenta,      
    urlcolor=blue,
    citecolor=blue,
}

\author{Can P. Ko\c{c}er}
\affiliation{Theory of Condensed Matter, Cavendish Laboratory, University of Cambridge, J. J. Thomson Avenue, Cambridge CB3 0HE, U.K.}
\author{Kent J. Griffith}
\affiliation{Department of Materials Science and Engineering, Northwestern University, Evanston, Illinois, 60208, USA}
\alsoaffiliation{Department of Chemistry, University of Cambridge, Lensfield Road, Cambridge CB2 1EW, U.K.}
\author{Clare P. Grey}
\affiliation{Department of Chemistry, University of Cambridge, Lensfield Road, Cambridge CB2 1EW, U.K.}
\author{Andrew J. Morris}
\affiliation{School of Metallurgy and Materials, University of Birmingham, Edgbaston, Birmingham B15 2TT, U.K.}
\email{a.j.morris.1@bham.ac.uk}

\title{{\large Cation Disorder and Lithium Insertion Mechanism of Wadsley--Roth Crystallographic Shear Phases from First Principles}}

\begin{document}

\begin{abstract}

\begin{figure}
\centering
\includegraphics[scale=0.24]{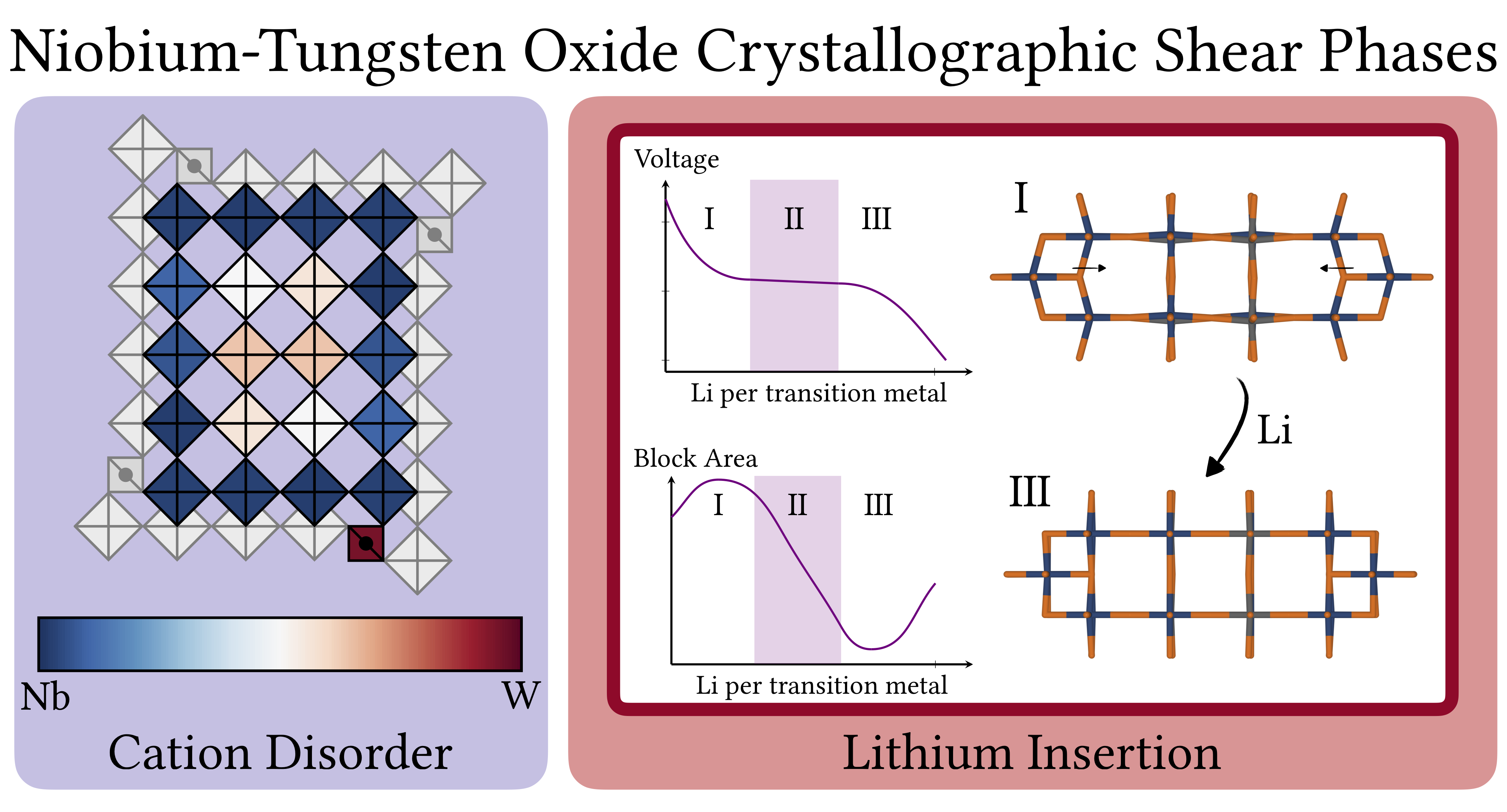}
\end{figure}
Wadsley--Roth crystallographic shear phases form a family of compounds that have attracted attention due to their excellent performance as lithium-ion battery electrodes. The complex crystallographic structure of these materials poses a challenge for first-principles computational modelling and hinders the understanding of their structural, electronic and dynamic properties. In this article, we study three different niobium-tungsten oxide crystallographic shear phases (\ce{Nb12WO33}, \ce{Nb14W3O44}, \ce{Nb16W5O55}) using an enumeration-based approach and first-principles density-functional theory calculations. We report common principles governing the cation disorder, lithium insertion mechanism, and electronic structure of these materials. Tungsten preferentially occupies tetrahedral and block-central sites within the block-type crystal structures, and the local structure of the materials depends on the cation configuration. The lithium insertion proceeds via a three-step mechanism, associated with an anisotropic evolution of the host lattice. Our calculations reveal an important connection between long-range and local structural changes: in the second step of the mechanism, the removal of local structural distortions leads to the contraction of the lattice along specific crystallographic directions, buffering the volume expansion of the material. Niobium-tungsten oxide shear structures host small amounts of localised electrons during initial lithium insertion due to the confining effect of the blocks, but quickly become metallic upon further lithiation. We argue that the combination of local, long-range, and electronic structural evolution over the course of lithiation is beneficial to the performance of these materials as battery electrodes. The mechanistic principles we establish arise from the compound-independent crystallographic shear structure, and are therefore likely to apply to niobium-titanium oxide or pure niobium oxide crystallographic shear phases.

\end{abstract}

\section{Introduction}

\begin{figure*}[htb]
    \centering
    \includegraphics[scale=0.31]{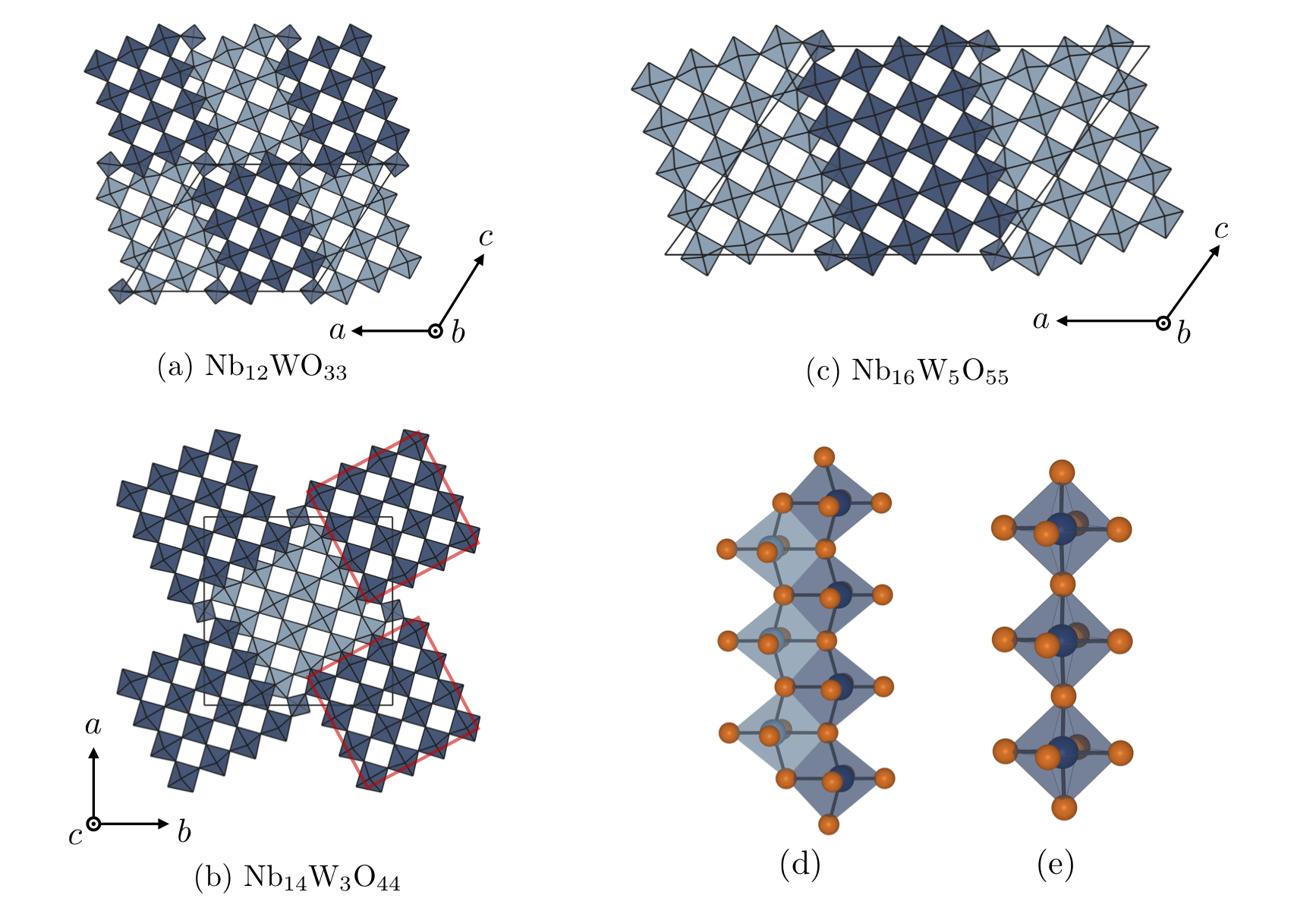}
    \caption{Crystal structures of Wadsley--Roth phases studied in this work: (a) \ce{Nb12WO33} (spacegroup $C2$), (b) \ce{Nb14W3O44} ($I4/m$), (c) \ce{Nb16W5O55} ($C2$). Light and dark coloured blocks are offset by half the lattice parameter perpendicular to the plane of the page. Note the increase in block size from a) \ce{Nb12WO33} ($3\times4$) to d) \ce{Nb16W5O55} ($4\times5$). The blocks are framed by crystallographic shear planes (edges of red squares in (b)), along which the metal-oxygen octahedra are strongly distorted (d). The octahedra in the block centre (e) are much less distorted. Transition metal atoms shown in blue and oxygen in orange.}
    \label{fig:xtalstrucs}
\end{figure*}

%{\bf [Li-ion batteries]}
There is a high demand for energy storage materials with improved performance in terms of energy and power density, cycle life, and safety. High-rate electrode materials specifically are needed to accelerate the adoption of electric vehicles by increasing power density and decreasing charging times. While strategies like nanostructuring have been used extensively to improve high-rate performance in materials like LTO\cite{chen2013} (\ce{Li4Ti5O12}), this has many drawbacks, including high cost, poor stability, and poor volumetric energy density\cite{wagemaker2013}. However, nanostructuring is not always necessary to obtain high rates. Recent work has shown that very high rates can be achieved in micrometre-sized particles of complex oxides of niobium (T-\ce{Nb2O5}\cite{griffith2016}), ternary Nb/W oxides (\ce{Nb16W5O55} and \ce{Nb18W16O93}~\cite{griffith2018}), and ternary Ti/Nb oxides (\ce{TiNb24O62}~\cite{griffith2017} and \ce{TiNb2O7}). In addition to the high-rate capability of these materials, their voltage range of +2.0 V to +1.0 V vs. Li$^+$/Li 
minimises electrolyte degradation and SEI formation, and avoids safety issues such as lithium dendrite formation.

%{\bf [Wadsley-Roth Crystallography and Crystal chemistry]}
Crystallographically, these complex oxides fall into two structural families: compounds with a tungsten bronze-type structure (T-\ce{Nb2O5}\cite{kato1975,griffith2016} and \ce{Nb18W16O93}\cite{griffith2018}), and Wadsley--Roth phases with block-type structures. The present work is concerned with the family of Wadsley--Roth phases, which encompasses a large number of crystallographically similar compounds in the \ce{Nb2O5}--\ce{WO3}~\cite{roth1965a} and \ce{Nb2O5}--\ce{TiO2}~\cite{wadsley1961} phase diagrams, in addition to pure \ce{Nb2O5}~\cite{kato1976} and \ce{Nb2O_{5-\delta}}~\cite{cava1991a} phases. The crystal structures of these compounds consist of blocks of corner-sharing octahedra of size $n\times m$, which are connected to each other by edge-sharing (Fig. \ref{fig:xtalstrucs}). The edge-sharing connections between the octahedra are present along so-called crystallographic shear planes, which frame the blocks. Perpendicular to the $n\times m$ plane the units connect infinitely (Fig. \ref{fig:xtalstrucs}d,e), and tetrahedral sites are present between the blocks in some structures to fill voids. Locally, the structures show strongly distorted octahedra due to a combination of electrostatic repulsion between cations and the second-order Jahn-Teller (SOJT) effect~\cite{kunz1995,bersuker2006}. \ce{NbO6} octahedra at the block periphery are more strongly distorted than those in the centre, resulting in zigzag-like patterns of metal cations along the crystallographic shear planes (Fig. \ref{fig:xtalstrucs}d). The block size depends in part on the oxygen-to-metal ratio of the compound; a higher number of oxygens per metal allows more corner-sharing connections between octahedra, and therefore larger blocks.

%{\bf [Wadsley-Roth phases as high-rate Li-ion battery anodes \& Previous work]}
Lithium insertion into Wadsley--Roth phases was first studied systematically by Cava et al. in 1983~\cite{cava1983}. The authors examined 12 different niobium oxide-based shear structures and showed that the crystallographic shear stabilises the structures against undesirable octahedral tilt distortions of the host framework, which had previously been observed in \ce{ReO3} \cite{cava1981}. The frustration of distortions allows lithium diffusion pathways to be kept open. Since the initial report by Cava et al., there have been articles detailing the electrochemical properties of many Wadsley--Roth phases, including \ce{TiNb2O7}~\cite{lu2011,guo2014}, \ce{Ti2Nb10O29}~\cite{cheng2014,wu2012}, \ce{TiNb24O62}~\cite{griffith2017}, \ce{Nb12WO33}~\cite{saritha2010,yan2017}, \ce{Nb14W3O44}~\cite{fuentes1997,yan2019}, \ce{Nb16W5O55}~\cite{griffith2018}, \ce{Nb12O29}~\cite{li2018,li2011}, H-\ce{Nb2O5}~\cite{griffith2016}, and \ce{PNb9O25}~\cite{patoux2002}. These studies have shown good performances of Wadsley--Roth phases as Li-ion battery electrodes, with a remarkable high-rate capability~\cite{griffith2018,guo2014}. Ultrafast lithium diffusion was recently observed in \ce{Nb16W5O55} with pulsed field gradient NMR spectroscopy and electrochemical techniques~\cite{griffith2018}. A strong similarity in the structural and phase evolution between different Wadsley--Roth phases has been noted~\cite{cava1983,griffith2018}. The phase evolution and voltage profile up to 1.5 Li/TM (Li per transition metal) can generally be divided into three regions; a first solid solution region with a sloping voltage profile is followed by a two-phase-like region where the voltage profile slope is flatter. Depending on the specific Wadsley--Roth phase, this second region of the voltage profile might be almost flat (as in H-\ce{Nb2O5}~\cite{griffith2016}), or have a small slope (\ce{Nb16W5O55}~\cite{griffith2018}). Beyond the two-phase-like region, another solid solution ensues. The similarity of their electrochemistry is highlighted by the fact that most articles reporting properties of a single Wadsley--Roth phase draw comparisons to other compounds of the family~\cite{griffith2017,griffith2018,patoux2002,cava1983,saritha2010}. Cation ordering preferences (such as in the Ti/Nb oxides~\cite{cheetham1973}) and electronic structure features~\cite{cava1991a,kocer2019} are also very similar.

%{\bf [Our work]}
Despite the rapidly growing number of experimental studies on Wadsley--Roth phases, reports of computational modelling are almost absent. First-principles modelling of Wadsley--Roth phases is both difficult and computationally expensive; the crystal structures are complex, have large unit cells with a multitude of lithium sites, and, in Nb/Ti and Nb/W oxides, feature inherent cation disorder. In this work, we study the cation disorder, lithium insertion mechanism, and electronic structure of three different Wadsley--Roth phases (\ce{Nb12WO33}, \ce{Nb14W3O44}, and \ce{Nb16W5O55}) using first-principles density-functional theory calculations. Their similarity in terms of both structure (cf. Fig.~\ref{fig:xtalstrucs}) and composition calls for a combined study. Building on our previous work on the electronic structure of \ce{Nb2O_{5-\delta}} crystallographic shear phases~\cite{kocer2019}, this study is motivated by the recent report of structural mechanisms in \ce{Li_xNb16W5O55}~\cite{griffith2018}, which we aim to understand from first principles.

The article is structured as follows. We begin by studying the Nb/W cation disorder using an enumeration approach. We establish cation ordering preferences and the lowest-energy cation configurations, and discover a variability of the local structure caused by the cation disorder. Next, we present a lithium insertion mechanism for \ce{Nb12WO33} in terms of the sequence of occupied lithium sites, the voltage profile, and the local and long-range structural evolution. We show that the mechanistic principles established for \ce{Nb12WO33} are transferable to \ce{Nb14W3O44} and \ce{Nb16W5O55}. In fact, \ce{Nb12WO33} and \ce{Nb14W3O44} can serve as model compounds to study the more complex \ce{Nb16W5O55}. After investigating the electronic structure of the materials over the course of lithium insertion, we go on to discuss common mechanistic principles for this structural family, and their implications for battery performance. We conclude by suggesting new directions for theory and experiment on structural, dynamic, and electrochemical properties of Wadsley--Roth phases.

\section{Methods}\label{Methods}

{\bf Structure enumeration.} Symmetrically distinct cation configurations of Nb/W within the (primitive, single block) unit cells of \ce{Nb14W3O44} and \ce{Nb16W5O55} were enumerated with a homemade program using established techniques~\cite{grau-crespo2007} based on a reduction of configurational space by the space group symmetry of a parent structure. Overall, 172 cation configurations were enumerated for \ce{Nb14W3O44}, and 45 for \ce{Nb16W5O55}. Further details can be found in the Supporting Information and the Results section.

The minority cation occupancy (i.e. tungsten occupancy) $P_{S}$ for site $S$ within the crystal structure was obtained according to

\begin{equation}\label{eq:Boltzmann}
	P_{S} = \frac{1}{Z} \sum_i \frac{N_{S,i}}{m_S}\ g_i e^{-\frac{E_i}{k_BT}}\,,
\end{equation}
where the symmetrically inequivalent cation configurations are labelled by $i$, and their degeneracy and energy above the ground state (per unit cell) are $g_i$ and $E_i$, respectively. $N_{S,i}$ denotes the number of positions of type $S$ that are occupied by tungsten in cation configuration $i$, and $m_S$ is the total number of positions of type $S$ within the unit cell. The partition function is given by $Z = \sum_i g_i e^{-\frac{E_i}{k_BT}}$. Equation \ref{eq:Boltzmann} can be understood as a thermodynamic average of the fraction of positions of type $S$ occupied by tungsten. The lowest energy cation configuration of each phase was used as a starting point to generate structural models of lithiated phases.

Structures of lithiated phases were generated by enumerating all possible lithium-vacancy configurations over sets of lithium sites in \ce{Nb12WO33} and \ce{Nb14W3O44}. The crystal symmetry was kept during this enumeration. Overall, 2048 structures were enumerated for \ce{Nb12WO33}, and 256 for \ce{Nb14W3O44}. Due to the much larger number of possible lithium sites in \ce{Nb16W5O55}, a full enumeration of lithium-vacancy configurations and subsequent DFT optimisation was computationally too expensive. Further details regarding the generation of lithiated structures can be found in the Results section.

{\bf Computational details.} All calculations were performed using the planewave pseudopotential DFT code CASTEP \cite{clark2005} (version 18.1). The gradient-corrected Perdew-Burke-Ernzerhof exchange-correlation functional for solids~\cite{perdew2008} (PBEsol) was used in the calculations presented in this work, unless otherwise specified. Many of the results we report are structural, and the PBEsol functional was therefore chosen because it provides better agreeement with experimental lattice parameters than PBE or LDA~\cite{perdew2008}. However, all of the results presented in this article show the same trends if computed with PBE instead. Structural optimisations were always performed in two steps: an initial relaxation using efficient parameters, followed by re-optimisation using very high accuracy parameters. For efficient parameters, core electrons were described using Vanderbilt ``ultrasoft'' pseudopotentials~\cite{vanderbilt1990}, generated using the `efficient' specifications listed in Table S1. These require smaller planewave kinetic energy cutoffs than the `high accuracy' ones. The planewave basis set was truncated at an energy cutoff of 400~eV, and integration over reciprocal space was performed using a Monkhorst-Pack grid~\cite{monkhorst1976} with a spacing finer than $2\pi \times 0.05$~\AA$^{-1}$. Higher accuracy was used to refine low-energy lithiated structures and all cation configurations. Harder, more transferable ultrasoft pseudopotentials were generated using the CASTEP 18.1 ``on-the-fly'' pseudopotential generator with the `high accuracy' specifications listed in Table S1. The planewave cutoff energy was set to 800~eV, and the Monkhorst-Pack grid spacing was chosen to be $2\pi \times 0.05$~\AA$^{-1}$ for calculations on pristine \ce{Nb12WO33}, \ce{Nb14W3O44} and \ce{Nb16W5O55} structures. For the lithiated phases, the Monkhorst-Pack grid spacing was set to $2\pi \times 0.03$~\AA$^{-1}$ due to their metallicity. Spin polarisation had a negligible effect on total energies, and structure optimisations using PBEsol were therefore performed without spin polarisation. Atomic positions and lattice parameters of all structures were optimised until the force on each atom was smaller than 0.01~eV/\AA, and the maximum displacement of any atom over two consecutive optimisation steps was smaller than $10^{-3}$~\AA.

DFT+$U$ calculations (following the method of Ref.~\cite{dudarev1998}) were performed to assess the impact of a change in the level of theory on thermodynamics and electronic structure. A value of $U = 4$ eV was chosen for the niobium and tungsten $d$-orbitals if not specified otherwise. This choice is in line with previous work~\cite{kocer2019} on niobium oxides. We note (and later demonstrate) that the results are mostly independent of the inclusion and exact value of the $U$ parameter.

{\bf Thermodynamics.} The thermodynamic phase stability of lithiated niobium-tungsten oxide phases was assessed by comparing the formation energy of different phases. For the pseudobinary phases considered in this work, a formation energy is defined as

\begin{equation}
    E_f = \frac{E\{\ce{Li_xY}\} - x E\{\mathrm{Li}\} - E\{\mathrm{Y}\}}{1+x}
\end{equation}
for Y = \ce{Nb12WO33}, \ce{Nb14W3O44}, or \ce{Nb16W5O55}. The formation energies were plotted as a function of the Li number fraction $c_{\mathrm{Li}} = \frac{\mathrm{x}}{1+\mathrm{x}}$. A pseudo-binary convex hull was constructed between the Y and Li end members at $(c_{\mathrm{Li}}, \mathrm{E_f})=(0,0);(1,0)$. Thermodynamically stable phases at $0$ K lie on the convex hull tieline.

%{\bf [Voltage profile]}
Voltages for transitions between phases lying on the convex hull were calculated from the DFT total energies. For two phases on the hull, $\mathrm{Li}_{\mathrm{x}_1}\mathrm{Y}$ and $\mathrm{Li}_{\mathrm{x}_2}\mathrm{Y}$, with $\mathrm{x}_2 > \mathrm{x}_1$, the voltage $V$ for a reaction

\begin{equation}
    \ce{Li_{x_1}Y} + (\mathrm{x}_2 - \mathrm{x}_1)\ \mathrm{Li} \to \ce{Li_{x_2}Y}
\end{equation}
is given by

\begin{equation}\label{eq:Voltage}
\begin{aligned}
    V ={} & -\frac{\Delta G}{x_2-x_1} \approx -\frac{\Delta E}{x_2 - x_1} \\
    = & -\frac{E(\mathrm{Li}_{x_2}\mathrm{Y}) - E(\mathrm{Li}_{x_1}\mathrm{Y})}{x_2-x_1} + E(\mathrm{Li})\,,
\end{aligned}
\end{equation}
where the Gibbs free energy is approximated by the internal energy, as the $pV$ and thermal contributions are small\cite{aydinol1997}.

{\bf Electronic structure and postprocessing.} Bandstructure calculations were performed for high-symmetry Brillouin zone directions according to those obtained from the SeeK-path package~\cite{hinuma2017}, which relies on spglib~\cite{togo2018}. A spacing between $\mathbf{k}$-points of $2\pi\times 0.025$ \si{\angstrom}$^{-1}$ was used for the bandstructures. Density of states calculations were performed with a grid spacing of $2\pi\times 0.01$ \si{\angstrom}$^{-1}$, and the results were postprocessed with the \texttt{OptaDOS} package~\cite{morris2014a} using the linear extrapolative scheme~\cite{pickard1999,pickard2000}. The \texttt{c2x}~\cite{rutter2018} utility and VESTA~\cite{momma2011} were used for visualisation of crystal structures and density data. Data analysis and visualisation was performed with the \texttt{matador}~\cite{evans} package.

\begin{figure*}[!htb]
	\centering
	\subfloat{\includegraphics[scale=0.165]{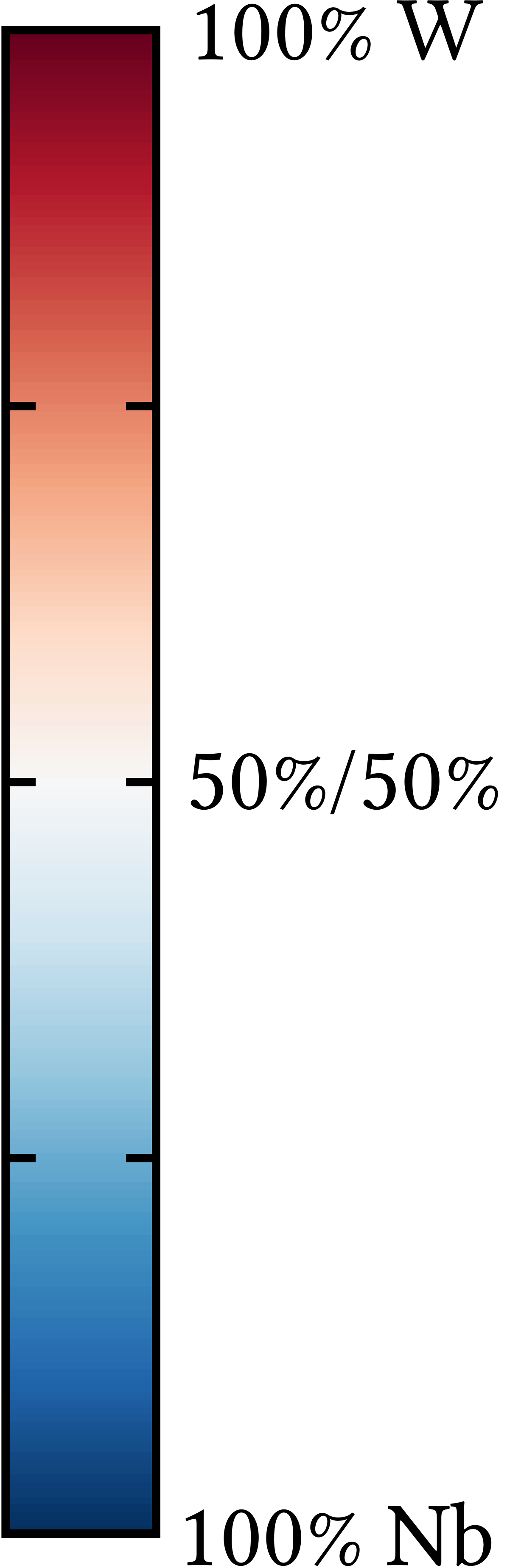}}\hspace{1.2cm}\setcounter{subfigure}{0}
	\subfloat[\ce{Nb14W3O44}]{\includegraphics[scale=0.16]{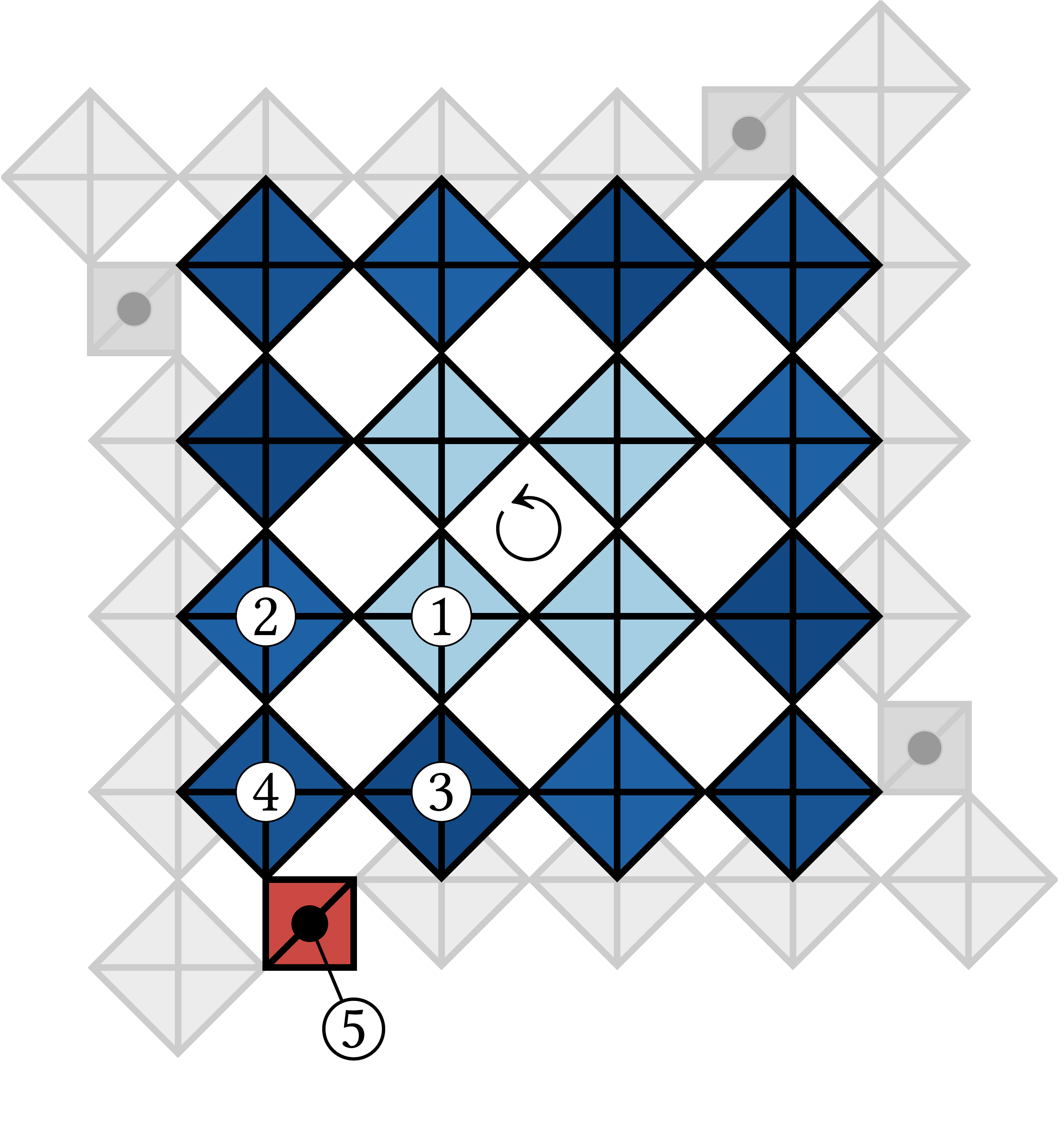}} \hspace{1.8cm}
	\subfloat[\ce{Nb16W5O55}]{\includegraphics[scale=0.16]{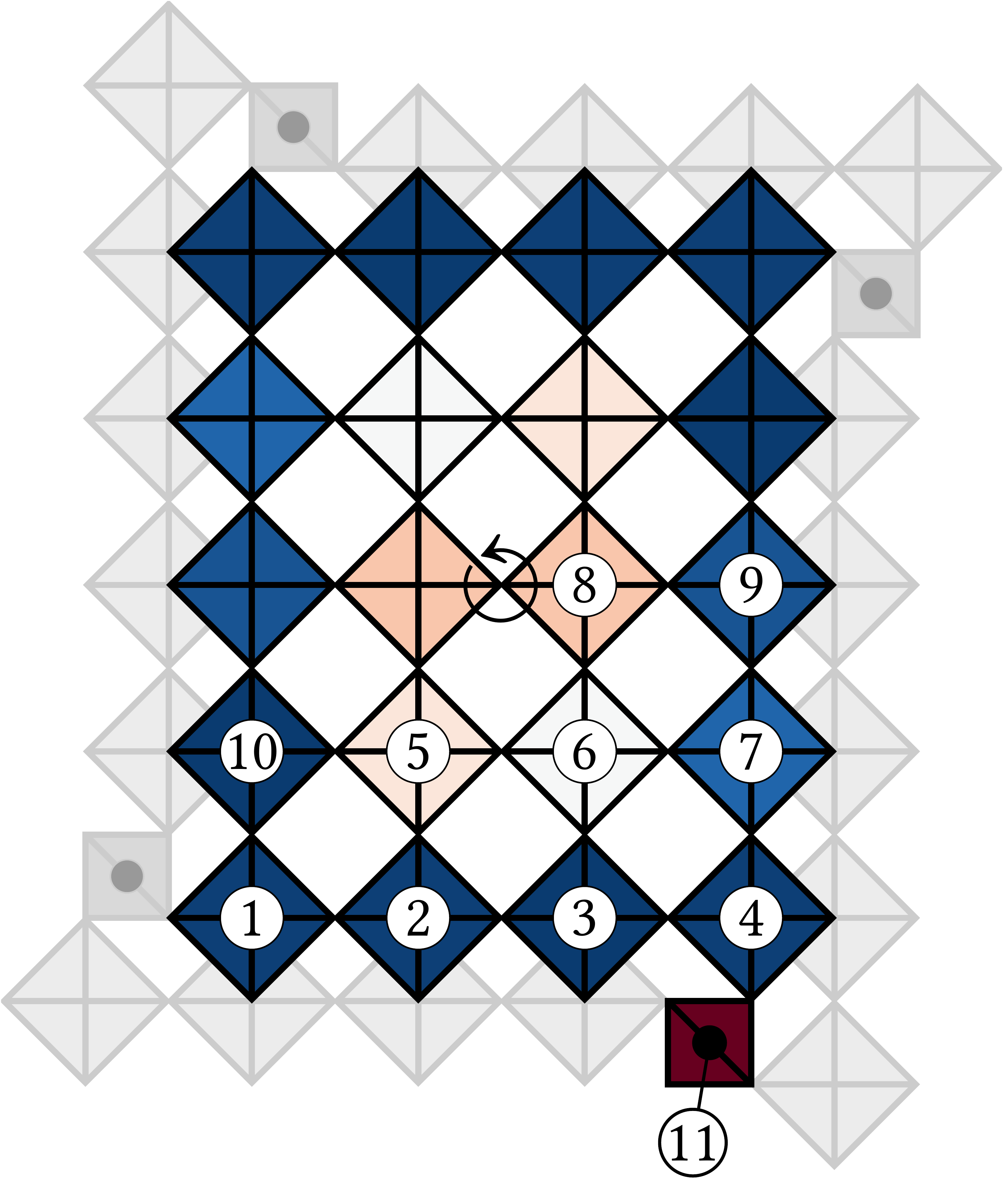}}
	\caption{Symmetrically inequivalent transition metal cation sites and their occupancies in (a) \ce{Nb14W3O44}, and (b) \ce{Nb16W5O55}. The labelling follows Cheetham and von Dreele~\cite{cheetham1983} for \ce{Nb14W3O44}, and Wadsley and Roth~\cite{roth1965b} for \ce{Nb16W5O55}. A temperature of 1200~$^\circ$C was used to determine the cation occupancies. The positions of axes of fourfold symmetry (\ce{Nb14W3O44}) and twofold symmetry (\ce{Nb16W5O55}) are indicated by circling arrows. In both structures, tungsten preferentially occupies the tetrahedral and block-central sites.}
	\label{fig:NbW_cationocc}
\end{figure*}

\begin{table*}[!htb]
    \centering
    \begin{tabular}{|c|c|c|c|c|}
    	\hline \hline
        Site & Expt.\cite{cheetham1983} & DFT (1050 $^{\circ}$C) & DFT (1200 $^{\circ}$C) & DFT (1350 $^{\circ}$C) \\ \hline \hline
        M1 & $0.39 \pm 0.04$ & 0.343 & 0.328 & 0.316 \\
        M2 & $0.23 \pm 0.07$ & 0.084 & 0.093 & 0.101 \\
        M3 & $0.00 \pm 0.06$ & 0.045 & 0.054 & 0.062 \\
        M4 & $0.00 \pm 0.05$ & 0.058 & 0.068 & 0.076 \\
        M5 (tet.) & $0.54 \pm 0.11$ & 0.877 & 0.830 & 0.782 \\
        \hline \hline
    \end{tabular}
    \caption{Tungsten occupancies on cation sites in \ce{Nb14W3O44}. All sites except M5 have a multiplicity of four. Taking into account the degeneracies, the number of tungsten atoms in a single block (Fig.~\ref{fig:NbW_cationocc}) is three, as required. The synthesis temperature is reported as 1350 $^{\circ}$C~\cite{roth1965a}, or 1050 $^{\circ}$C~\cite{cheetham1983}. Note that the refinement of fractional occupancies reported in Ref.~\cite{cheetham1983} was performed in $I4/m$, while the DFT predictions are for $I\bar{4}$. The multiplicity of the tetrahedral site is different in these two spacegroups, and the experimental occupancy has been adjusted accordingly. The experimental data~\cite{cheetham1983} includes estimated standard deviations.}
    \label{tab:Nb14W3O44_cationocc}
\end{table*}

\section{Results}

\subsection{Cation Disorder}

%{\bf [Occupancy preferences]}
Neutron diffraction studies have established that the cation distribution in block-type structures is disordered but not random~\cite{cheetham1973,cheetham1983,griffith2017}. Some amount of disorder is also suggested by single crystal X-ray diffraction studies~\cite{wadsley1961,roth1965b}. Labelling conventions for the cation sites in the crystal structures are shown in Figure~\ref{fig:NbW_cationocc}, and abide by literature conventions as much as possible. To derive fractional occupancies for the tungsten cations in \ce{Nb14W3O44} and \ce{Nb16W5O55} we apply a Boltzmann distribution (Eqn.~\ref{eq:Boltzmann}) using the DFT total energies of the symmetrically inequivalent cation configurations. The results are listed in Tables \ref{tab:Nb14W3O44_cationocc} and S2 for temperatures of 1050--1350~$^{\circ}$C, which corresponds to the range of synthesis and annealing temperatures~\cite{roth1965a,cheetham1983,griffith2018}. Cation occupancies in \ce{Nb14W3O44} and \ce{Nb16W5O55} at 1200~$^\circ$C are presented in Fig.~\ref{fig:NbW_cationocc} using a colormap. Plots of the tungsten occupancies for an extended temperature range are available in the Supporting Information (Figs. S1, S2).

%{\bf [\ce{Nb14W3O44}, DFT]}
If the cation distribution in \ce{Nb14W3O44} was completely random, each site would have a tungsten occupancy of $\frac{3}{17}\approx0.176$. Instead, tungsten is predicted to favour the M5 tetrahedral position and the M1 block-center position (Table~\ref{tab:Nb14W3O44_cationocc}). The preferential occupancy of tungsten on the purely corner-shared M1 position is expected; the metal-metal distances are larger in the block center, and the occupation of these sites by the more highly charged tungsten cations (assuming W$^{6+}$ vs. Nb$^{5+}$) reduces the overall electrostatic repulsion. Preferential occupation of tungsten on the tetrahedral site is due to the shorter M-O distances, which, together with the higher charge of the tungsten cations, lead to better covalency and stronger bonds. In fact, the 15 lowest energy structures generated by enumeration and DFT optimisation all have tungsten on the tetrahedral site. The two lowest energy cation configurations both have the tetrahedral site occupied by tungsten, in addition to two M1 sites. The lowest energy configuration has spacegroup $C2$, whereas the second lowest configuration has spacegroup $P1$ (+123 meV/f.u. above groundstate). The highest energy structure lies +1.29 eV/f.u. above the ground state.

%{\bf [Expt]}
The cation ordering in \ce{Nb14W3O44} has previously been investigated by Cheetham and Allen using neutron powder diffraction~\cite{cheetham1983}. DFT-derived fractional occupancies are in reasonable agreement with experiment (Table~\ref{tab:Nb14W3O44_cationocc}). The overall sequence of site occupancy preferences is the same. The occupancy of the tetrahedral site M5 is predicted to be larger, while the occupancy of M2 is predicted to be much smaller. Those two site occupancies also have the largest estimated experimental uncertainty (Table~\ref{tab:Nb14W3O44_cationocc}). Given the very similar local structures of M2, M3, and M4, the large occupancy of M2 as compared to M3 and M4 seems inconsistent. Determining occupancies in these large and complex structures is difficult, particularly when the neutron scattering lengths are not very different ($7.054$ and $4.86\times 10^{-15}$ m for Nb and W, respectively)~\cite{sears1992}.  We suggest that the cation distribution should be revisited, perhaps with a joint X-ray/neutron study, to help constrain the occupancies.

%{\bf [\ce{Nb12WO33}]}
X-ray diffraction studies suggest that the tungsten atom in \ce{Nb12WO33} is ordered on the tetrahedral site~\cite{roth1965b}. An enumeration within the primitive unit cell of \ce{Nb12WO33} produces only 7 structures, for the 7 symmetrically inequivalent sites. Placing the tungsten atom on the tetrahedral site results in the lowest energy structure. The second lowest energy structure with tungsten in the block-center lies +364 meV/f.u. above the ground state, suggesting a strong preference for the tetrahedral site even compared to the block-center position.

%{\bf [\ce{Nb16W5O55}]}
Experimental data regarding the cation ordering in \ce{Nb16W5O55} is not available. However, the structure of \ce{Nb16W5O55} is very similar to that of \ce{Nb14W3O44}, with only one additional row of octahedra within each block. For our calculations, the tetrahedral site has been fully occupied by tungsten given the preference of tungsten for the tetrahedral site in \ce{Nb14W3O44} and \ce{Nb12WO33}. We have also constrained ourselves to configurations in space group $C2$. The more highly charged tungsten cations again prefer to occupy the purely corner-shared octahedral positions in the block middle of \ce{Nb16W5O55}; occupancies of sites M5, M6, and M8 are by far the largest (Fig.~\ref{fig:NbW_cationocc}, Table S2).  The lowest energy cation configuration for \ce{Nb16W5O55} has tungsten on sites M8 and M5, while the second and third lowest energy configurations have tungsten on sites M8 and M6 (+11 meV/f.u. vs. groundstate) and M5 and M6 (+147 meV/f.u. vs. groundstate). The highest energy cation configuration lies +2.27 eV/f.u. above the groundstate.

%{\bf [Caveats]}
There are several effects that are not taken into account by the DFT prediction; (1) the modelling necessarily assumes that the material is in thermal equilibrium, but depending on synthesis temperature and annealing time, the kinetics of solid state diffusion might play a role in determining the site occupancies, (2) only single-block cation configurations were studied, limiting the length scale of interactions, (3) at the high synthesis temperature of the metal oxide, temperature effects such as volume expansion, harmonic or even anharmonic vibrations certainly play a role and the DFT energy is a good, but limited substitute for the full free energy. Nevertheless, the lowest energy single-block cation configurations are the best choice to use in modelling the lithiation mechanism, and are shown explicitly in Fig. S3. We include crystallographic information files (CIF) for all PBEsol-optimised symmetrically inequivalent cation configurations of \ce{Nb14W3O44} and \ce{Nb16W5O55} in the Supporting Information of this article, in addition to a table of their space groups, relative energies, and degeneracies.

%{\bf [Local structure variability]}
The individual cation configurations deviate from the idealised parent crystal structure by different amounts. For both \ce{Nb14W3O44} and \ce{Nb16W5O55}, the distributions of lattice parameters and unit cell volumes of the cation configurations show a spread of 1-2 \% around the mean. In addition to slight differences in lattice parameters, the \ce{MO6} octahedra of both \ce{Nb14W3O44} and \ce{Nb16W5O55} exhibit different distortions depending on the cation configuration. To analyse these distortions, we introduce three distortion measures: a dimensionless bond angle variance $\Delta(\theta_{\mathrm{oct}})$, the quadratic elongation $\lambda_{\mathrm{oct}}$, and an off-centering distance $d_{\mathrm{oct}}$. The bond angle variance and quadratic elongation are commonly used distortion measures~\cite{robinson1971} implemented, for example, in VESTA~\cite{momma2011}. The $\Delta(\theta_{\mathrm{oct}})$ measure is defined as the bond angle variance divided by the square of the mean to make the quantity dimensionless:

\begin{equation}\label{eq:Delta_oct}
\Delta(\theta_{\mathrm{oct}}) = \frac{1}{12} \sum_{i=1}^{12} \bigg[\frac{\theta_i - \langle\theta_i\rangle}{\langle\theta_i\rangle}\bigg]^2\,,
\end{equation}
where the 12 O-M-O angles are denoted by~$\theta_i$. Note that only angles which are 90$^\circ$ in an ideal octahedron are included.

\begin{figure}[!htb]
	\centering
	\includegraphics[scale=0.575]{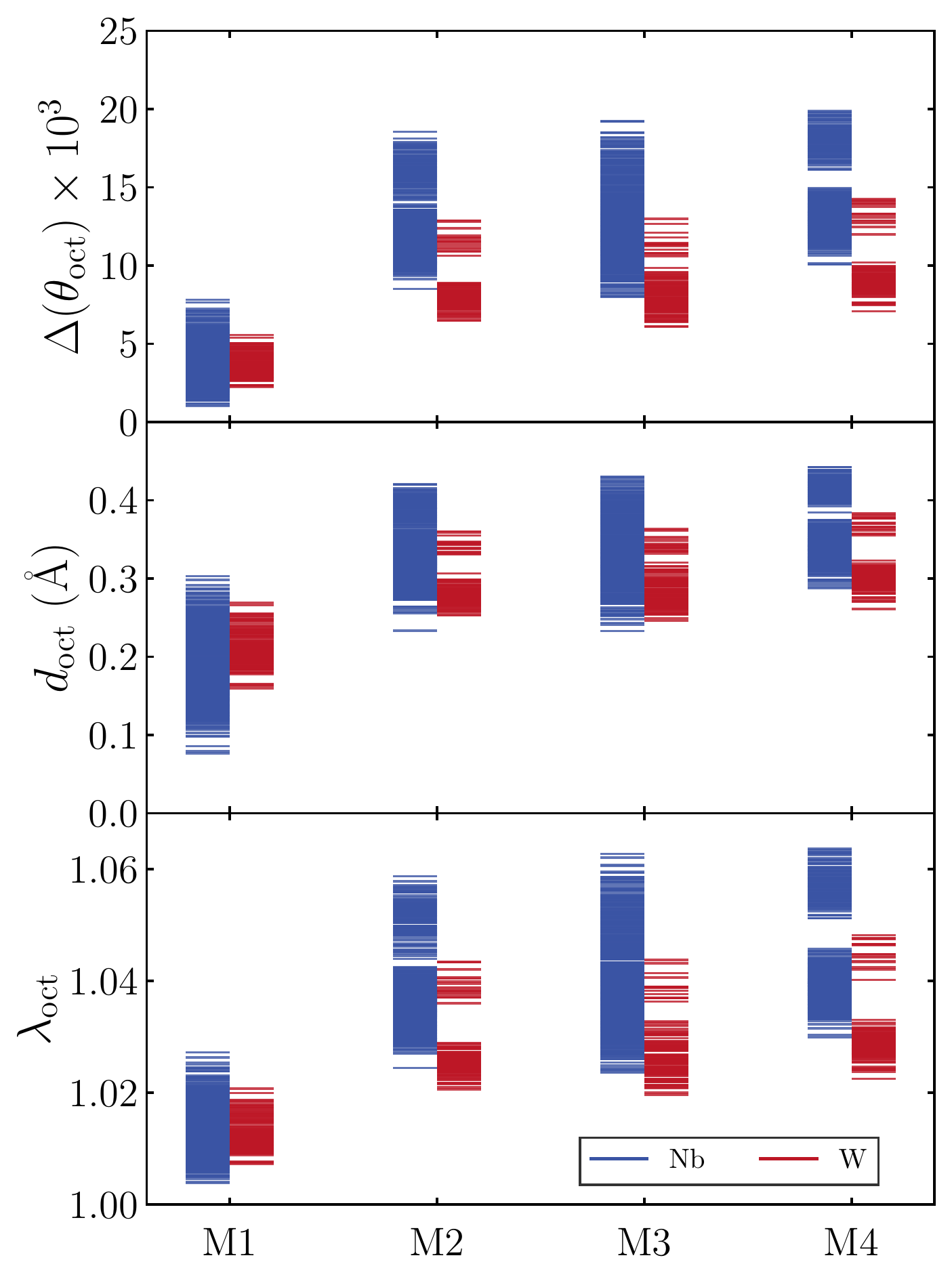}
	\caption{Distortion measures for octahedral positions M1--M4 (cf. Fig.~\ref{fig:NbW_cationocc}) for all 172 cation configurations of \ce{Nb14W3O44}. The block-central M1 octahedra are more symmetric than the peripheral M2--M4 octahedra. All sites show a significant spread in their octahedral distortion measures.}
	\label{fig:Nb14W3O44_catconfig_distortions}
\end{figure}

The quadratic elongation $\lambda_{\mathrm{oct}}$ is defined as 

\begin{equation}\label{eq:lambda_oct}
\lambda_{\mathrm{oct}} = \frac{1}{6} \sum_i^6 \Big(\frac{l_i}{l_0}\Big)^2\,,
\end{equation}
where $l_i$ are the M-O bond lengths, and $l_0$ is the M-O bond length for an octahedron with $O_h$ symmetry whose volume is equal to that of the distorted octahedron~\cite{robinson1971}. The off-centering distance is defined as the distance between the center of the \ce{O6} polyhedron and the metal position

\begin{equation}
d_{\mathrm{oct}} = \bigg\lVert\mathbf{r}_{\mathrm{M}} - \sum_{i=1}^{6} \frac{\mathbf{r}_{\mathrm{O},i}}{6} \bigg\rVert\,,
\end{equation}
where $\mathbf{r}_M$ is the metal position and $\mathbf{r}_{\mathrm{O},i}$ are the oxygen positions. Both $\Delta(\theta_{\mathrm{oct}})$ and $d_{\mathrm{oct}}$ are zero for an ideal octahedron, and $\lambda_{\mathrm{oct}}$ is one. The three distortion measures are plotted in Fig.~\ref{fig:Nb14W3O44_catconfig_distortions} for the M1--M4 sites in \ce{Nb14W3O44} for all 172 cation configurations, but we note that not all configurations will contribute equally due to their different Boltzmann weight.

The M1 block-center octahedron is, on average, less distorted than the the block-peripheral M2, M3, and M4 octahedra. However, all octahedral positions show a significant spread in their distortion measures, indicating a dependence of the local structure on the cation configuration. To put these results into context, we note that quadratic elongation measures for octahedra in inorganic compounds fall in the range of 1.00--1.07~\cite{robinson1971}. \ce{Nb14W3O44} exhibits this entire range of distortions if all transition metal sites and cation configurations are considered together. The off-center distances show a spread of approximately 0.15--0.2 \AA. Given the convergence tolerance of $10^{-3}$ \AA~for the DFT geometry optimisation, this indicates a significant static disorder in the atomic positions. Similar results are obtained for \ce{Nb16W5O55}, also showing weaker distortions for the block-central sites (M5, M6, M8) and a significant spread in the distortion measures for all transition metal octahedra in the structure (Fig. S4). Overall, these results indicate a variability of the local structure at the unit cell level in mixed-metal shear phases that is not captured by a single cation configuration. Each cation configuration has a different set of cation-cation neighbour patterns, which can cause different local distortion directions and strengths. In this study, only cation configurations within the primitive unit cell have been considered. Effects on a longer range can be important, and would lead to a more continuous variation of the local structure. For example, there are two sets within the distortion measures for tungsten on the M2 site (Fig.~\ref{fig:Nb14W3O44_catconfig_distortions}), separated by a gap. The more distorted set corresponds to \ce{WO6} octahedra edge-sharing with two other \ce{WO6} octahedra along the crystallographic shear plane, while the less distorted set corresponds to \ce{WO6} edge-sharing with two \ce{NbO6} octahedra. Configurations within a supercell along the $c$ direction (cf. Fig.~\ref{fig:xtalstrucs}) would include \ce{WO6} octahedra sharing edges with one \ce{NbO6} and one \ce{WO6} octahedron, and likely close the gap.

Both niobium and tungsten are generally classified as intermediate SOJT distorters within the group of $d^0$ cations~\cite{ok2006}. In \ce{Nb14W3O44}, niobium and tungsten show very similar distortion strengths on the M1 positions, while the distortion for tungsten seems to be weaker for sites M2--M4. Given the local structure variability in Nb/W oxide shear structures, it is very likely that the Ti/Nb structures show the same properties, since $d^0$ titanium is also classified as an intermediate distorter. Stronger distortions are generally exhibited by molybdenum, while zirconium shows only very weak distortions~\cite{ok2006}. It would be interesting to examine the effect of Mo/Zr doping on the local structure in shear phases.

\begin{figure*}[!htb]
    \centering
    \includegraphics[scale=0.22]{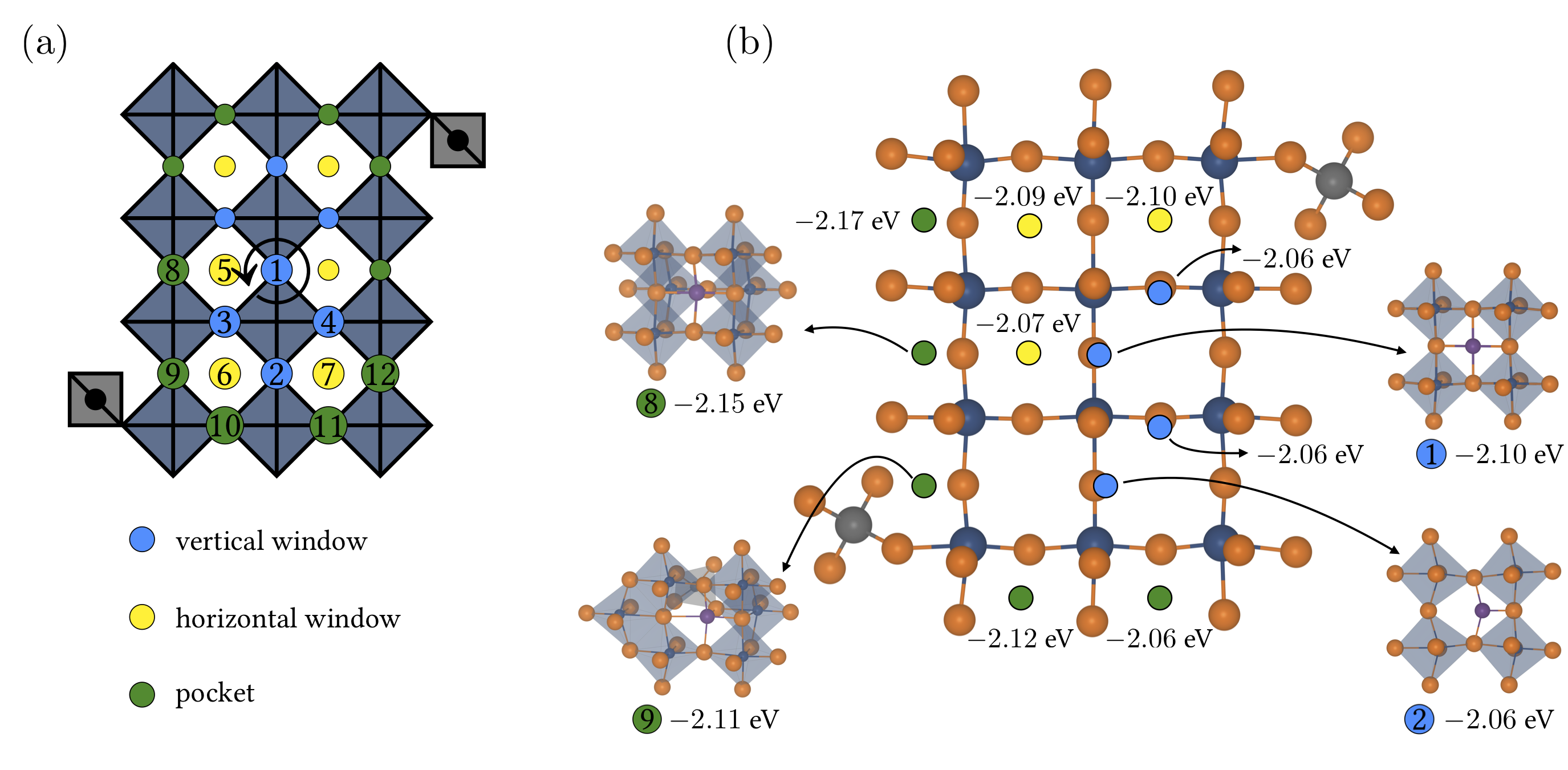}
    \caption{(a) Types of lithium sites present in \ce{Nb12WO33}. Window positions are fourfold coordinated by oxygens, pocket positions fivefold. The circling arrow marks the twofold rotation axis of the crystal structure. This symmetry element is kept for the enumeration of lithiated structures. (b) Local structure of lithium sites and site energies in \ce{Nb12WO33}. Only one of each pair of equivalent sites is shown. Insertion into fivefold coordinated sites is energetically more favourable. The vertical window positions next to the crystallographic shear planes (sites 2, 3, 4) are too large for fourfold coordination of lithium. Niobium shown in dark blue, lithium in purple, and oxygen in orange.}
    \label{fig:Nb12WO33singlesite}
\end{figure*}

\subsection{Lithium Insertion Mechanism}

\subsubsection{\ce{Nb12WO33}}

%{\bf [Single lithium sites, Enumeration]}
Lithium sites in block-type structures divide into three sets; fivefold coordinated `pocket' sites at the edge of the block, fourfold coordinated horizontal `window' positions, and fourfold coordinated vertical `window' positions (Fig.~\ref{fig:Nb12WO33singlesite}a). These sites have been deduced by neutron diffraction studies for lithiated block-type structures \ce{TiNb2O7} and H-\ce{Nb2O5}~\cite{catti2013,catti2015}. We will assume and verify the presence of these sites for \ce{Nb12WO33}. The lithium site energies and local structures in \ce{Nb12WO33} are shown in Fig. \ref{fig:Nb12WO33singlesite}b. Site energies and structures were obtained by placing a single lithium atom into a ($1\times 2\times 1$) supercell of \ce{Nb12WO33} (cf. Fig.~\ref{fig:xtalstrucs}) and optimising the structure. The site energies $E_{f,i}$ were calculated as

\begin{equation}
	E_{f,i} = E_{i} - E_{\mathrm{SC}} - E(\mathrm{Li})\,,
\end{equation}
where $E_{i}$ is the energy of the supercell with a lithium atom placed at site $i$, $E_{SC}$ is the energy of the supercell, and $E(\mathrm{Li})$ is the energy of bulk lithium. A comparison of the site energies shows that the insertion into fivefold coordinated sites is energetically more favourable. Horizontal window positions have a symmetric arrangement of oxygen atoms, while vertical window positions and some of the pocket sites are less symmetric. In the horizontal window position, the lithium ion sits slightly above the plane formed by the four oxygen atoms.  The vertical window positions (sites 2, 3, and 4) are too large for fourfold coordination of lithium by the oxygen atoms, and insertion into these sites is energetically the least favourable. The resulting threefold coordinated lithium ion moves far off the plane formed by the oxygens. The single site energies of around $-2.1$ eV agree well with the starting point of the voltage profile at 2 V vs. Li$^+$/Li~\cite{yan2017}.

\begin{figure}[!htb]
	\centering
	\includegraphics[scale=0.46]{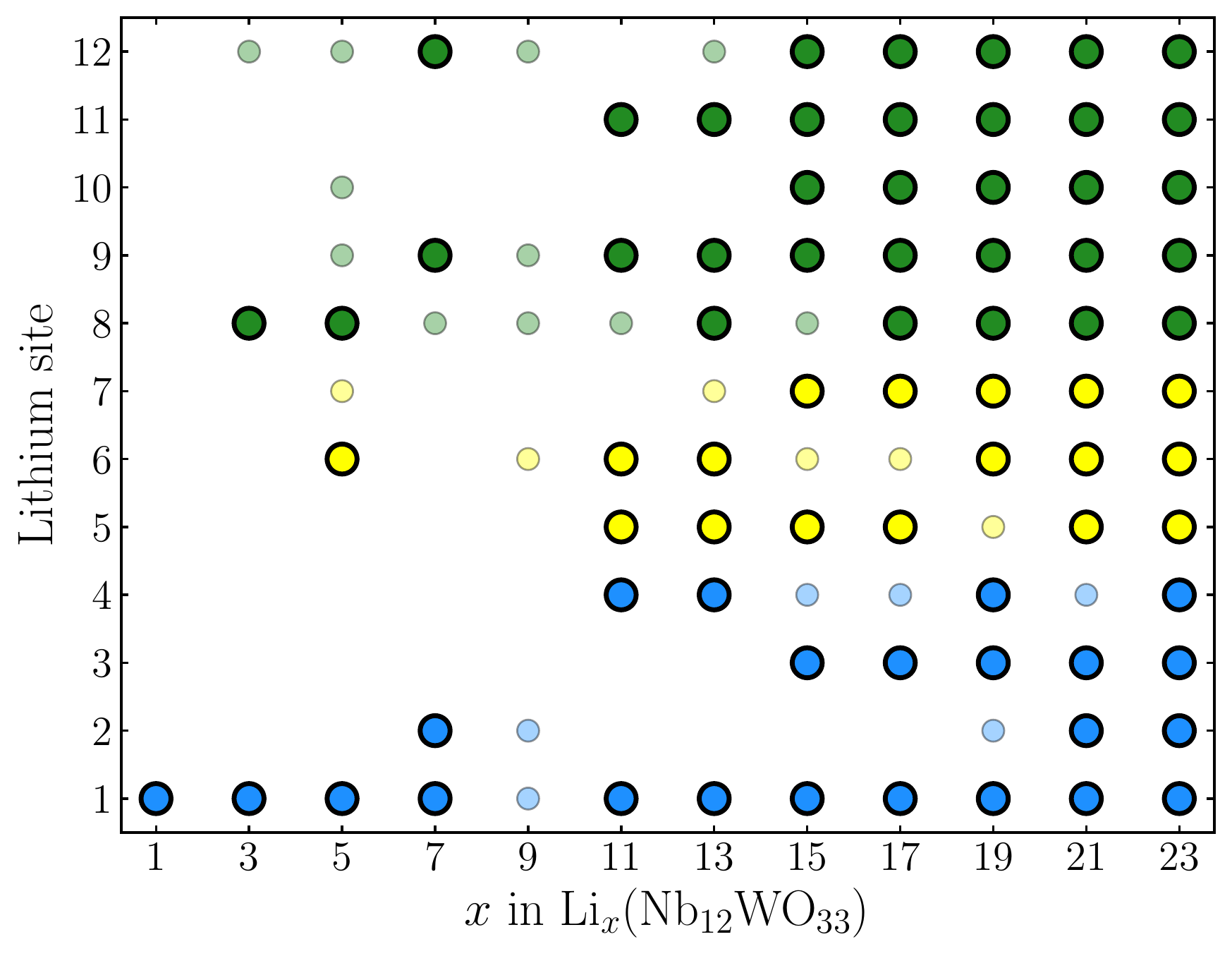}
	\caption{Occupation of lithium sites for each sampled stoichiometry. Lithium sites labelled according to Fig.~\ref{fig:Nb12WO33singlesite}. Bold dots correspond to sites occupied in the structure on the convex hull tieline, smaller dots mark sites that are occupied in structures up to 200 meV/f.u. above the convex hull tieline. There is no simple sequence of lithium site occupation.}
	\label{fig:Nb12WO33_siteocc_seq}
\end{figure}

In order to simulate the lithium insertion into \ce{Nb12WO33} over the entire range of lithium content, lithiated structures \ce{Li_xNb12WO33} were generated by enumerating all possible lithium-vacancy configurations for the sites shown in Fig.~\ref{fig:Nb12WO33singlesite}. The special position in the center of the block (site 1) was fixed to be occupied. Using the remaining 11 independent sites for \ce{Nb12WO33}, $2^{11}=2048$ lithiated structures result, for stoichiometries of \ce{Li_xNb12WO33} with $x$ ranging from 1 to 23 in steps of 2. This enumeration produces `snapshots' of the structure and energetics of \ce{Li_xNb12WO33} at specific stoichiometries. The convex hull of the lowest energy \ce{Li_xNb12WO33} structures (Fig. S5) shows stable or nearly stable phases for each of the stoichiometries examined, indicating that no extended two-phase regions occur. To reliably capture the lithium insertion mechanism, it is useful to include metastable structures (i.e. up to a certain cutoff energy above the convex hull tieline) in the analysis. These metastable structures could be accessed at finite temperatures. If only thermodynamically stable structures are considered, there is no simple sequence of occupation of lithium sites (Fig.~\ref{fig:Nb12WO33_siteocc_seq}), although there is a slight initial preference for occupation of fivefold coordinated sites and undistorted fourfold sites, especially if metastable structures are included. Both site energies (Fig.~\ref{fig:Nb12WO33singlesite}) and Li-Li interactions are important for determining the lithium insertion sequence.

\begin{figure}[!htb]
    \centering
    \includegraphics[scale=0.45]{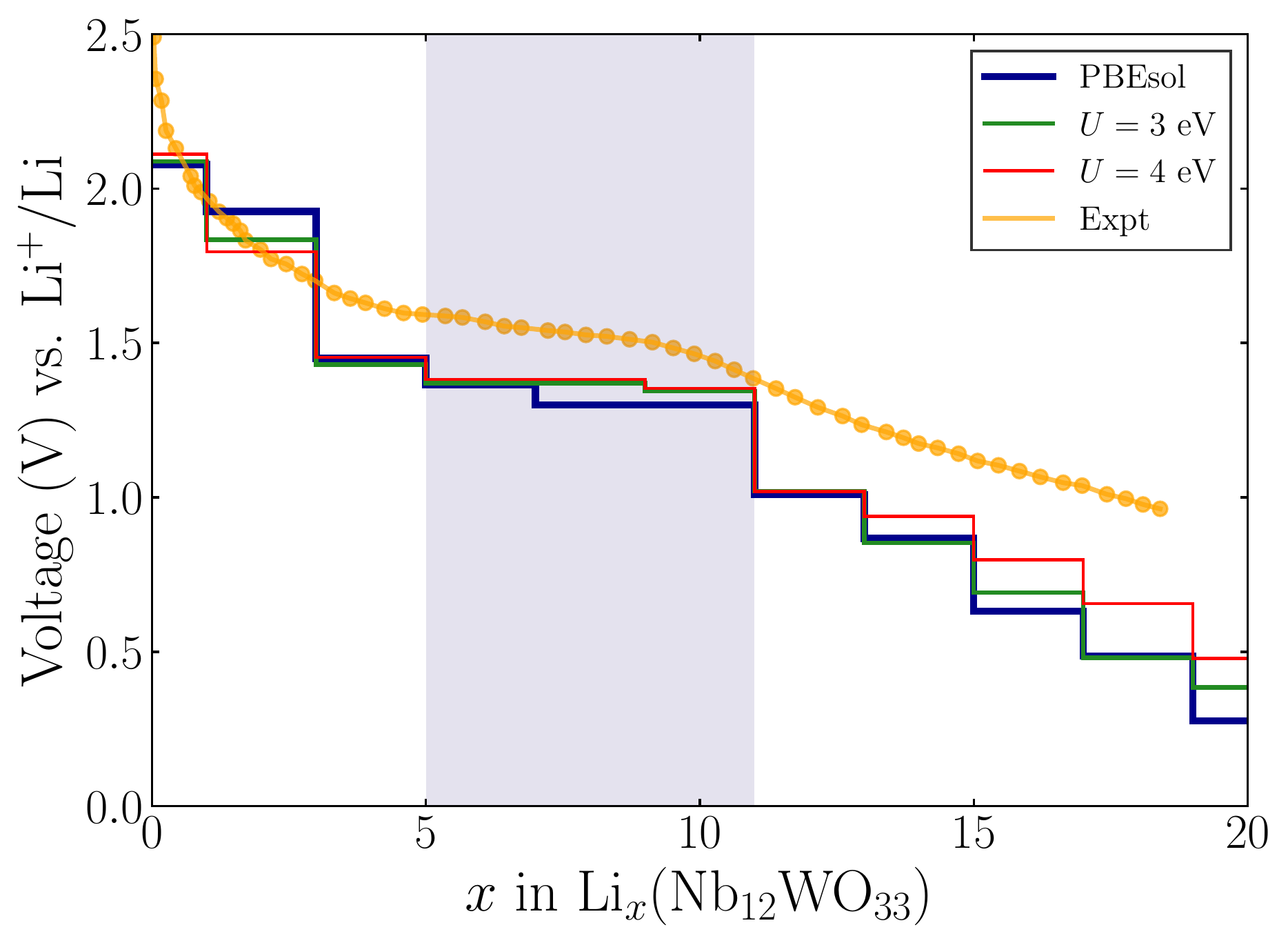}
    \caption{Experimental voltage profile (orange, digitised from Ref.~\cite{yan2017}) compared to DFT predictions: PBEsol (blue), PBEsol+$U$ for $U=3$ eV (green), and $U=4$ eV (red). The predicted voltage profiles are composed of steps due to the discrete sampling of stoichiometries, and are in qualitative agreement with the experimental profile.}
    \label{fig:Nb12WO33HullVoltage}
\end{figure}

%{\bf [Voltage profile]}
A comparison of the experimental~\cite{yan2017} and DFT-predicted voltage profiles (calculated with Eqn.~\ref{eq:Voltage}) at the GGA and GGA+$U$ levels of theory is shown in Figure~\ref{fig:Nb12WO33HullVoltage}. The DFT-predicted voltage profiles are necessarily composed of abrupt step changes due to the discrete number of stoichiometries, and only qualitative comparisons between experimental and DFT-predicted voltage profiles should be made. We also note that the experimental voltage profile has not explicitly been recorded under equilibrium conditions.

Compared to the experiment, PBEsol slightly underestimates the average insertion voltage; the average experimental voltage up to 1 Li/TM is 1.65 V, whereas PBEsol predicts 1.44 V. The average insertion voltages evaluated with PBE and LDA are 1.30 V and 1.70 V, respectively. We note that the inclusion of a $U$ value for the niobium $4d$ orbitals has a minor effect on the average insertion voltage; for both $U=3$ eV and $U=4$ eV, the average insertion voltage up to 1 Li/TM is 1.45 V. It is well known that GGA functionals underestimate lithium insertion voltages of transition metal oxides, but this can be corrected for late first-row elements (Fe/Mn/Co/Ni) by DFT+$U$ methods~\cite{urban2016}. The case of niobium oxides seems to be closer to that of $d^0$ titanium oxides, in that the use of DFT+$U$ is ineffective~\cite{dalton2012} (cf. Supplementary Methods). In addition it is unclear what the value of $U$ should be for this case; the electronic structure and chemical bonding will change as a function of lithium concentrations, possibly requiring different $U$ values at different points to be described accurately. However, total energies (and therefore phase stability) for sets of structures with different $U$ values cannot be compared. Since the difference between GGA and GGA+$U$ results is small, we will continue with a GGA treatment and defer discussion of the electronic structure to a later section. We note that while hybrid functionals like HSE06 are able to provide better agreement with experimental voltages, their use is computationally more expensive and errors of $\pm 0.2$~V are still common~\cite{urban2016}.

While the average insertion voltage is underestimated, the shape of the DFT-predicted profiles does show similarity to the experimental one; there seems to be a region with rather flat slope between $x=3$ and $x=11$, which matches the flatter second region of the experimental profile. Despite the shallow gradient of the electrochemical profile, this region does not correspond to a true two-phase region. The similarity between the experiment and DFT prediction is present for both the PBEsol and PBEsol+$U$ results, and becomes clearer if the predicted profiles are shifted upwards by the difference in the average insertion voltage, corresponding to an adjustment of the Li chemical potential (Fig. S6).

\begin{figure*}[!htb]
    \centering
    \subfloat[]{\includegraphics[scale=0.6]{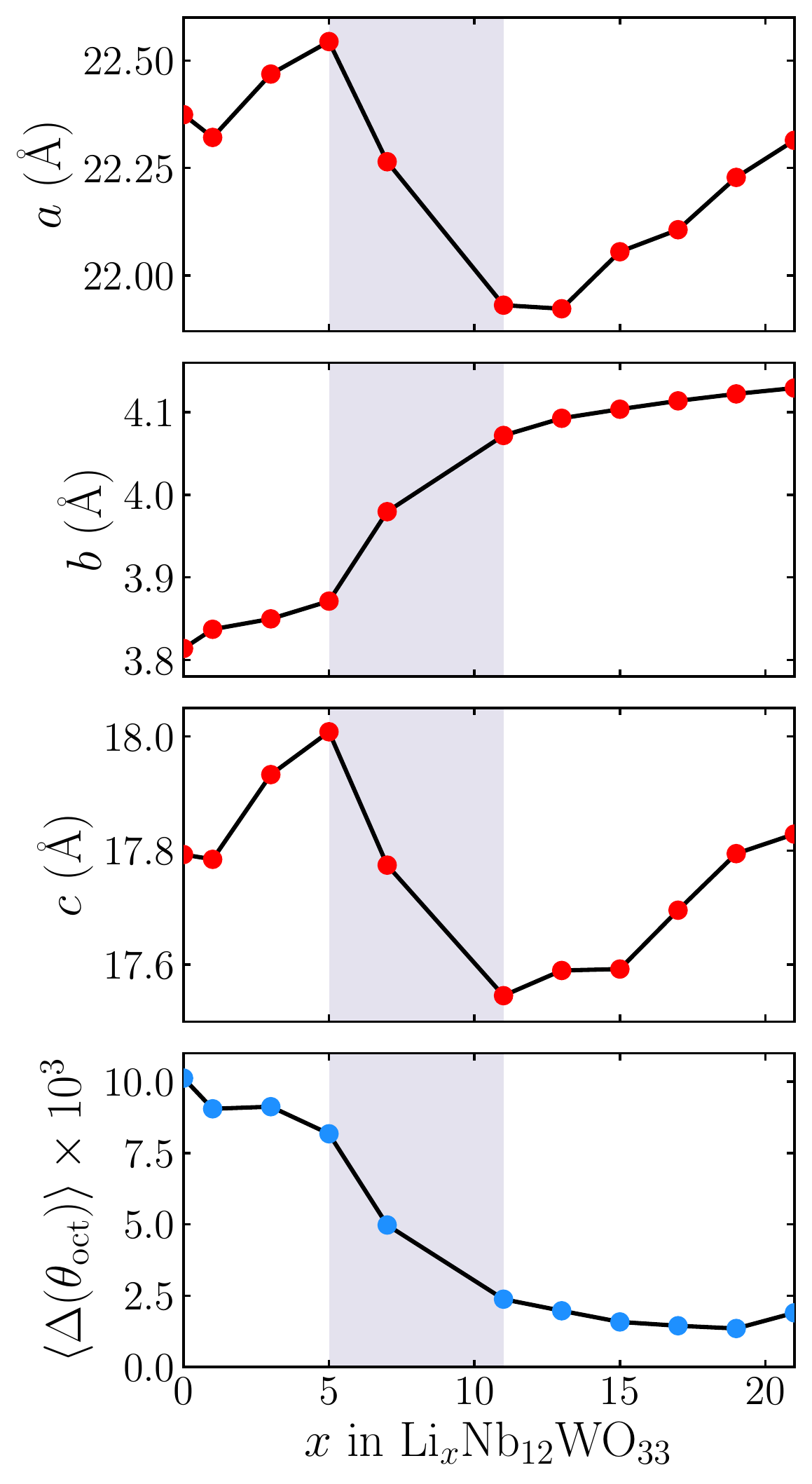}}\hspace{1.5cm}
    \subfloat[]{\includegraphics[scale=0.46]{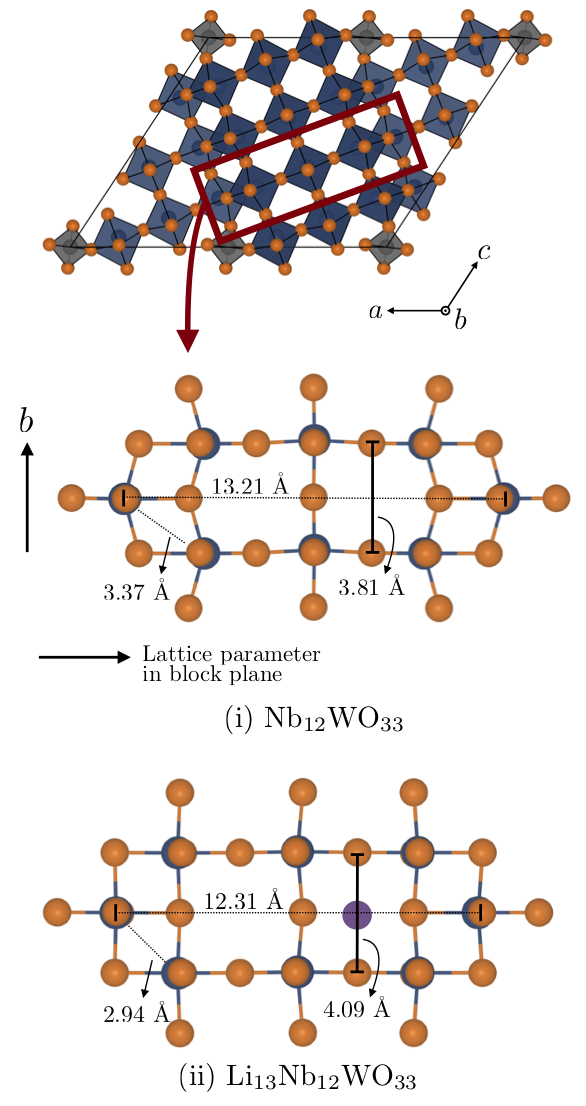}}
    \caption{Structural evolution of \ce{Li_xNb12WO33} as a function of lithium content $x$. (a) The lattice parameters evolve anisotropically; $b$ expands over the entire $x$ range, while $a$ and $c$ first expand until $x=5$, contract, and then expand again beyond $x=13$. The average octahedral distortion $\langle\Delta(\theta_{\mathrm{oct}})\rangle$ decreases, with most of the decrease between $x=5$ and $x=11$. (b) The local structure in (i) \ce{Nb12WO33} and (ii) \ce{Li13Nb12WO33} along the second row of octahedra in the $3\times 4$ block. Niobium in dark blue, oxygen in orange, and lithium in purple. The interatomic distances demonstrate 1) an expansion perpendicular to the block plane, 2) a contraction within the block plane, and 3) a decrease of Nb-Nb distances along the shear planes. Compared to \ce{Nb12WO33}, the \ce{NbO6} octahedra in \ce{Li13Nb12WO33} are more symmetric, corresponding to a smaller distortion measure $\langle \Delta(\theta_{\mathrm{oct}})\rangle$.}
    \label{fig:Nb12WO33_StrucEvol}
\end{figure*}

%{\bf [Lattice Evolution]}
The evolution of the lattice parameters of \ce{Nb12WO33} as a function of lithium content is anisotropic (Fig. \ref{fig:Nb12WO33_StrucEvol}a). Lattice parameter $b$, which is perpendicular to the plane of the block, expands, and most of the expansion takes place between $x=5$ and $x=11$. Lattice parameters $a$ and $c$ first expand until $x=5$, and then contract to a minimum at $x=11$ that lies almost $0.3$ \si{\angstrom} below the lattice parameters of the pristine structure. For $x > 11$, $a$ and $c$ expand again. The lattice contraction occurs in the same region as the flatter part of the voltage profile (shaded blue in Figs.~\ref{fig:Nb12WO33HullVoltage}, \ref{fig:Nb12WO33_StrucEvol}a). The same evolution of the lattice parameters is also observed when phases up to 200 meV/f.u. above the convex hull tieline are included in the analysis (Fig. S7). These metastable structures might be formed during cycling, or be partially accessible due to finite temperature effects. However, the same lattice evolution would result.

%{\bf [Changes in local structure]}
Over the course of lithium insertion, the transition-metal oxygen octahedra become progressively more symmetric, as shown by the evolution of the average distortion measure $\langle \Delta(\theta_{\mathrm{oct}})\rangle$ (Fig.~\ref{fig:Nb12WO33_StrucEvol}a), obtained according to

\begin{equation}\label{eq:Delta_oct_avg}
\langle\Delta(\theta_{\mathrm{oct}})\rangle = \frac{1}{N_{\mathrm{oct}}} \sum_{j=1}^{N_{\mathrm{oct}}} \Delta_j(\theta_{\mathrm{oct}})\,.
\end{equation}
Compared to the pristine \ce{Nb12WO33}, the distortions in both the \ce{MO6} octahedra and the lithium sites are largely removed in \ce{Li13Nb12WO33} (Fig.~\ref{fig:Nb12WO33_StrucEvol}b). The evolution in the lattice parameters and the local structure is closely linked. Over the course of lithium insertion, the blocks of octahedra in \ce{Nb12WO33} first expand and then contract within the $ac$ plane (Fig. \ref{fig:Nb12WO33_StrucEvol}). Perpendicular to the $ac$ plane, they expand monotonically. An expansion is expected for lithium insertion, as an increase of the number of atoms within the same volume should lead to an increase thereof. The decrease in the lattice parameters within the block is associated with the \ce{MO6} octahedra symmetrisation. As the apical oxygens of the octahedra along the shear planes are pulled towards the block center, the lattice shrinks  within the block plane (Fig.~\ref{fig:Nb12WO33_StrucEvol}b). The block height expands from 3.81 \si{\angstrom} to 4.09 \si{\angstrom}, and the Nb-Nb distance along the shear plane decreases by over 0.4 \si{\angstrom}.

The structural changes are closely connected to the occupation of specific lithium sites; the thermodynamically stable phases of \ce{Li_xNb12WO33} (Fig.~\ref{fig:Nb12WO33_siteocc_seq}) show occupation of undistorted sites (1, 6, and 8) for $x \leq 5$. For $x\geq 7$, vertical window positions that were previously highly distorted are occupied, and the distortions in both lithium sites and octahedra start to be removed.

Based on the predicted voltage profile, lattice evolution, and local structure changes, the overall phase evolution of \ce{Nb12WO33} through three regions can be rationalised. Taken together, and compared to previous experiments, these results suggest two solid solution regions, with a two-phase-like region in between. The two-phase-like region is marked by a block-plane contraction and a removal of distortions in the transition metal-oxygen octahedra.

\subsubsection{\ce{Nb14W3O44} and \ce{Nb16W5O55}}

Following on from \ce{Nb12WO33}, we now demonstrate that very similar lithium insertion mechanisms apply to \ce{Nb14W3O44} and \ce{Nb16W5O55}.

\begin{figure}[!htb]
	\centering
	\includegraphics[scale=0.375]{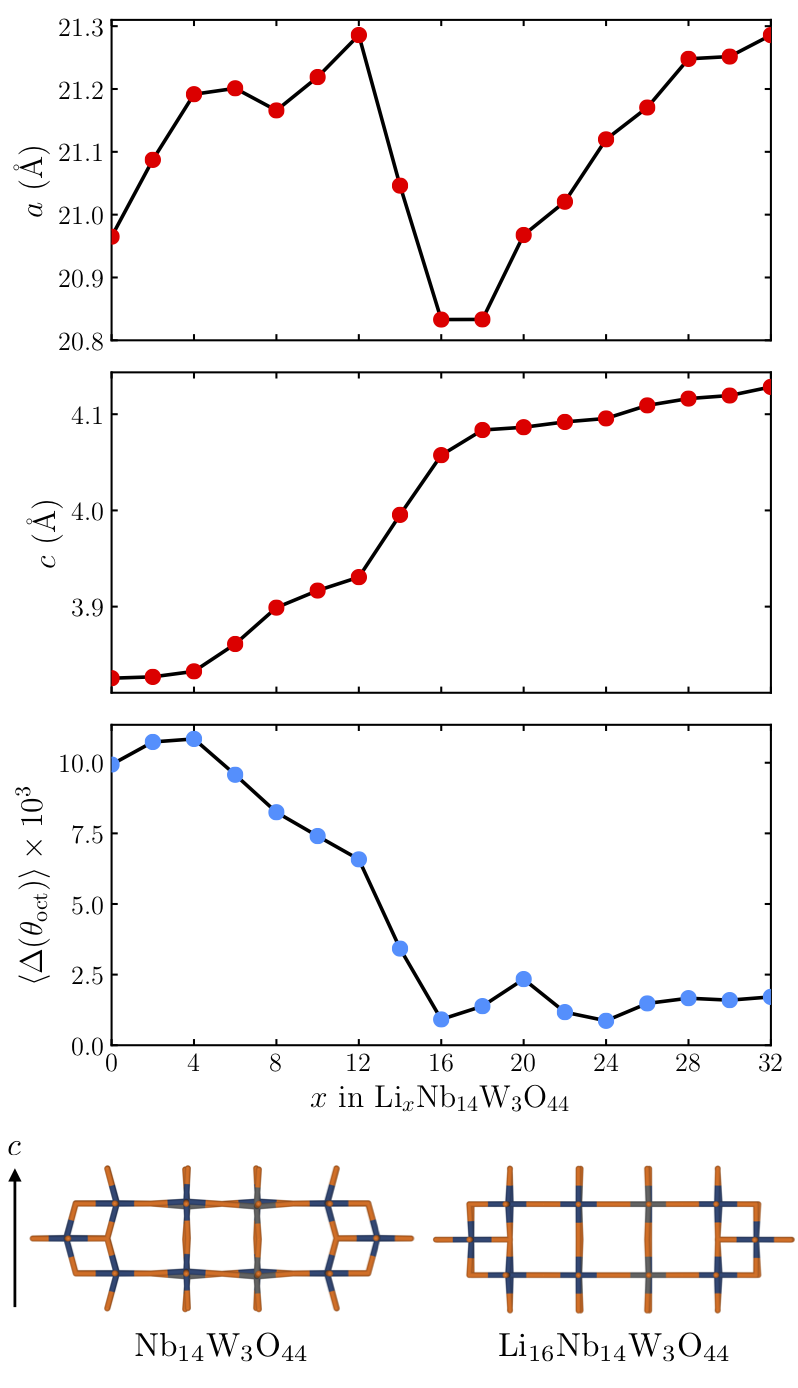}
	\caption{Local and long-range structural evolution of \ce{Nb14W3O44} during lithium insertion. The anisotropic lattice evolution and the removal of the octahedral distortions ($\langle\Delta(\theta_{\mathrm{oct}})\rangle$) strongly resembles \ce{Nb12WO33}. (cf. Fig.~\ref{fig:Nb12WO33_StrucEvol}). Compared to \ce{Nb14W3O44}, the transition metal-oxygen framework (bottom) for the fully lithiated \ce{Li16Nb14W3O44} structure shows significantly weaker octahedral distortions. Lithium ions have been omitted in \ce{Li16Nb14W3O44} for clarity. The removal of the distortions leads to a contraction of the lattice parameters within the block plane (perpendicular to $c$).}
	\label{fig:Nb14W3O44_StrucEvol}
\end{figure}

\begin{figure}[!htb]
	\centering
	\includegraphics[scale=0.32]{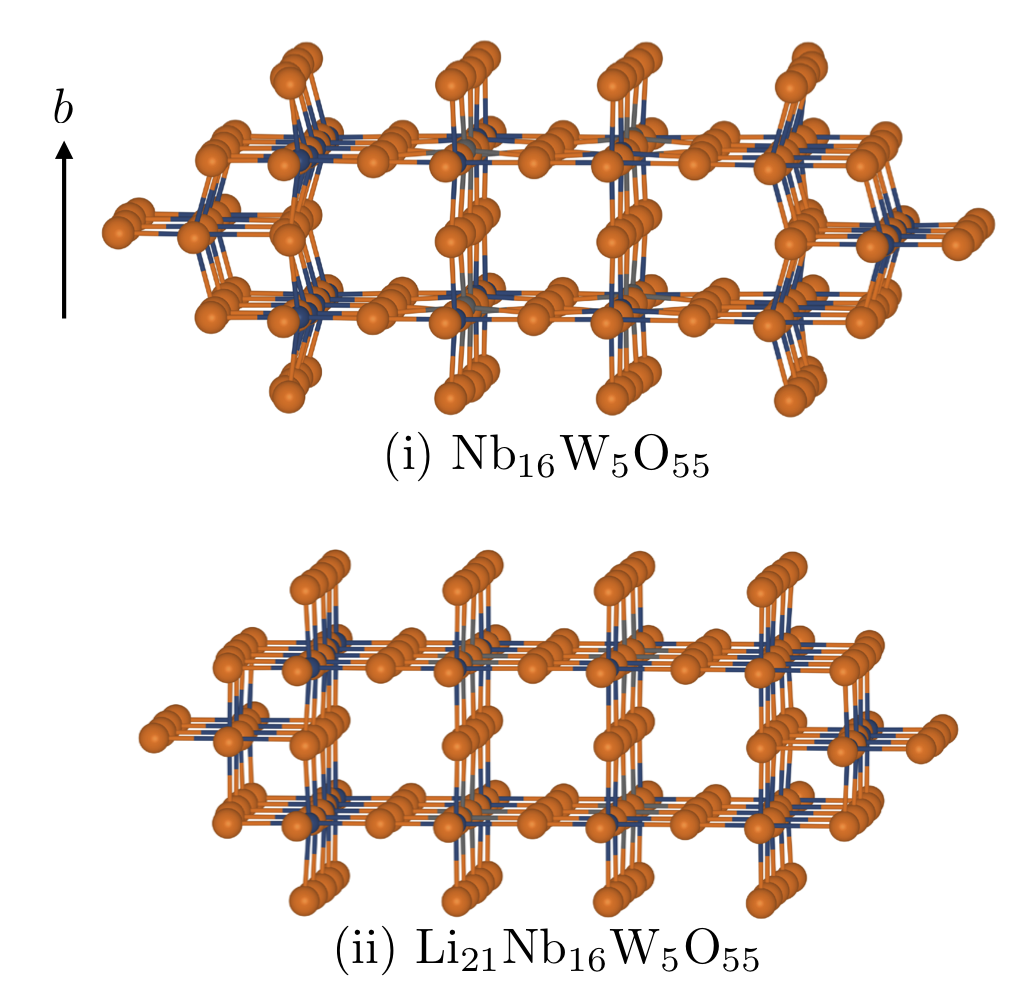}
	\caption{Structure of the transition-metal oxygen framework in pristine and fully lithiated \ce{Nb16W5O55}. Lithium ions have been omitted in the lithiated structure for clarity. The transition metal-oxygen framework in \ce{Li21Nb16W5O55} shows significantly more symmetric \ce{MO6} octahedra. The removal of the distortions leads to a contraction of the lattice parameters within the block plane (perpendicular to $b$).} 
	\label{fig:WNBO_phases}
\end{figure}

%{\bf [Lithium sites]}
The crystal structures of \ce{Nb12WO33}, \ce{Nb14W3O44}, and \ce{Nb16W5O55} are all are based on the block principle and feature the same local structural distortions (cf. Fig.~\ref{fig:xtalstrucs}). This similarity leads to the presence of the same types of lithium environments in all three structures. The classification into pocket and window sites in \ce{Nb14W3O44} and \ce{Nb16W5O55} follows the same principles as for \ce{Nb12WO33} (Fig. S8). Notably, the vertical window positions next to the crystallographic shear planes (sites 3 and 5 in \ce{Nb14W3O44}, and sites G, H, K in \ce{Nb16W5O55}, Fig. S9) are strongly distorted due to the zigzag patterns of the octahedra (cf. Fig. \ref{fig:xtalstrucs}). Lithium site energies for \ce{Nb14W3O44} are in the range of -2.0 eV to -2.2 eV, while the site energies for \ce{Nb16W5O55} are slightly lower (-2.2 eV to -2.4 eV), due to the higher concentration of tungsten (cf. Table S3). Insertion into fivefold coordinated sites is energetically favoured.

%{\bf [Structure generation, Convex Hull]}
The enumeration for \ce{Li_xNb14W3O44} was performed in the same way as for \ce{Li_xNb12WO33}. The special position in the center of the block was fixed to be unoccupied. Given the remaining 8 lithium sites, $2^8=256$ structures were enumerated. The number of lithiated structures generated by enumeration for \ce{Nb14W3O44} is much smaller compared to \ce{Nb12WO33}. Structures with lithium content between those covered by enumeration were `interpolated' by using the low-energy enumerated structures as a starting point. For example, candidate structures of \ce{Li6Nb14W3O44} were generated by filling half of the lithium sites occupied in \ce{Li4Nb14W3O44} and \ce{Li8Nb14W3O44}. Overall, the sampling of lithiated structures for \ce{Nb14W3O44} is coarser than for \ce{Nb12WO33}, due to the higher computational cost of optimising the lithium configurations in a larger unit cell. A convex hull of the lowest energy \ce{Li_xNb14W3O44} phases is available in the Supporting Information (Fig. S10), and shows thermodynamically stable phases at every sampled stoichiometry. A full enumeration of lithium-vacancy configurations in \ce{Nb16W5O55} is not possible. The primitive unit cell contains 22 independent lithium sites, resulting in $2^{22}=4194304$ possible lithium-vacancy configurations.

%{\bf [Structural evolution]}
The structural evolution of \ce{Nb14W3O44} over the course of lithium insertion (Fig.~\ref{fig:Nb14W3O44_StrucEvol}) bears a strong resemblance to that of \ce{Nb12WO33} (cf. Fig.~\ref{fig:Nb12WO33_StrucEvol}). Lattice parameter $c$, perpendicular to the block plane, expands monotonically, with most of the expansion taking place between $x=12$ and $x=16$ (Fig. \ref{fig:Nb14W3O44_StrucEvol}). The parameter $a$ first increases, then shrinks below its initial value with a minimum at $x=16$. Another expansion for $x>18$ follows. Note that lattice parameter $a$ (which is equal to $b$ in the $I\bar{4}$ spacegroup of \ce{Nb14W3O44}) was extracted as $a=\sqrt{V/c}$ (cf. Fig.~\ref{fig:xtalstrucs}). The same trend in the evolution of the lattice parameters is also observed when phases up to 100 meV/f.u. above the convex hull tieline are included in the analysis (Fig. S11). The distortions of the \ce{MO6} octahedra are removed as demonstrated by the decrease in the $\langle\Delta(\theta_{\mathrm{oct}})\rangle$ measure (Eqn.~\ref{eq:Delta_oct_avg}). The contraction and distortion removal is associated with occupation of the distorted vertical window positions, in direct analogy to \ce{Nb12WO33}. However, the extent of the structural regions differs between \ce{Nb12WO33} and \ce{Nb14W3O44}. In \ce{Nb12WO33}, the maximum expansion of the $a$ and $c$ parameters occurs at 0.4 Li/TM, while in \ce{Nb14W3O44} it occurs at 0.71 Li/TM. The contraction region is also wider in \ce{Nb12WO33}; it spans from 0.38 Li/TM to 1.0 Li/TM, while in \ce{Nb14W3O44}, the contraction occurs from 0.71 Li/TM to 1.06 Li/TM. It is difficult to decide whether this is a physically significant difference, or simply due to the smaller number of lithium configurations that were sampled for \ce{Nb14W3O44} as compared to \ce{Nb12WO33}.

%{\bf [Lithiated structures]}
Lithium insertion into \ce{Nb14W3O44} initially proceeds via occupation of sites 1, 4, and 8 (cf. Fig. S8), but overall there is no simple sequence for the filling of lithium sites. The lowest energy structures for each stoichiometry are available as crystallographic information files (CIF) in the Supporting Information.

%{\bf [Link]}
In complete analogy to \ce{Nb12WO33}, the local and long-range structural changes in \ce{Nb14W3O44} are linked. The removal of the distortions of the \ce{MO6} octahedra along the shear planes pulls the blocks closer together (Fig.~\ref{fig:Nb14W3O44_StrucEvol}). As a result, the lattice parameter in the block plane, $a$, decreases.

%{\bf [\ce{Nb16W5O55}]}
While we cannot perform a thorough sampling of lithium-vacancy configurations for \ce{Nb16W5O55}, the strong structural similarity between these three niobium-tungsten oxides suggests that the same trend of lattice and local structural evolution will apply to \ce{Nb16W5O55}. As a proof-of-principle, we have produced a structural model for \ce{Li21Nb16W5O55} by occupying sites E, I, J, L, N, M, and G (cf. Fig. S8), which is shown in Fig.~\ref{fig:WNBO_phases}. Compared to the pristine structure, the lithiated structure shows a contraction in the block plane ($a=29.54\ \mathrm{\AA}$ vs. $a=29.34$ \AA, $c=23.10\ \mathrm{\AA}$  vs. $c=22.95$ \AA, for \ce{Nb16W5O55} and \ce{Li21Nb16W5O55} respectively), and an expansion perpendicular to the block plane ($b=3.81\ \mathrm{\AA}$  vs. $b=4.06$ \AA), in good quantitative agreement with experimental findings~\cite{griffith2018}. The octahedral distortion measure $\langle\Delta(\theta_{\mathrm{oct}})\rangle$ decreases from $10.25\times 10^{-3}$ for \ce{Nb16W5O55} to $0.86\times 10^{-3}$ for \ce{Li21Nb16W5O55}. Clearly, lithium insertion causes the same overall structural changes in all three niobium-tungsten oxides \ce{Nb12WO33}, \ce{Nb14W3O44}, and \ce{Nb16W5O55}.

\subsection{Electronic Structure of Lithiated Phases}

%{\bf [Importance]}
In this section, we briefly present key electronic structure features of niobium-tungsten oxide shear phases. The electronic structure of the shear structures determines their electronic conductivity, which is important for high-rate battery performance. Additionally, the results presented here serve to explain the mixed-metal redox process and to justify the level of theory used in this study. We will focus on \ce{Nb14W3O44}, but the results are transferable to \ce{Nb12WO33} and \ce{Nb16W5O55}.

\begin{figure}[!htb]
	\centering
	\includegraphics[scale=0.45]{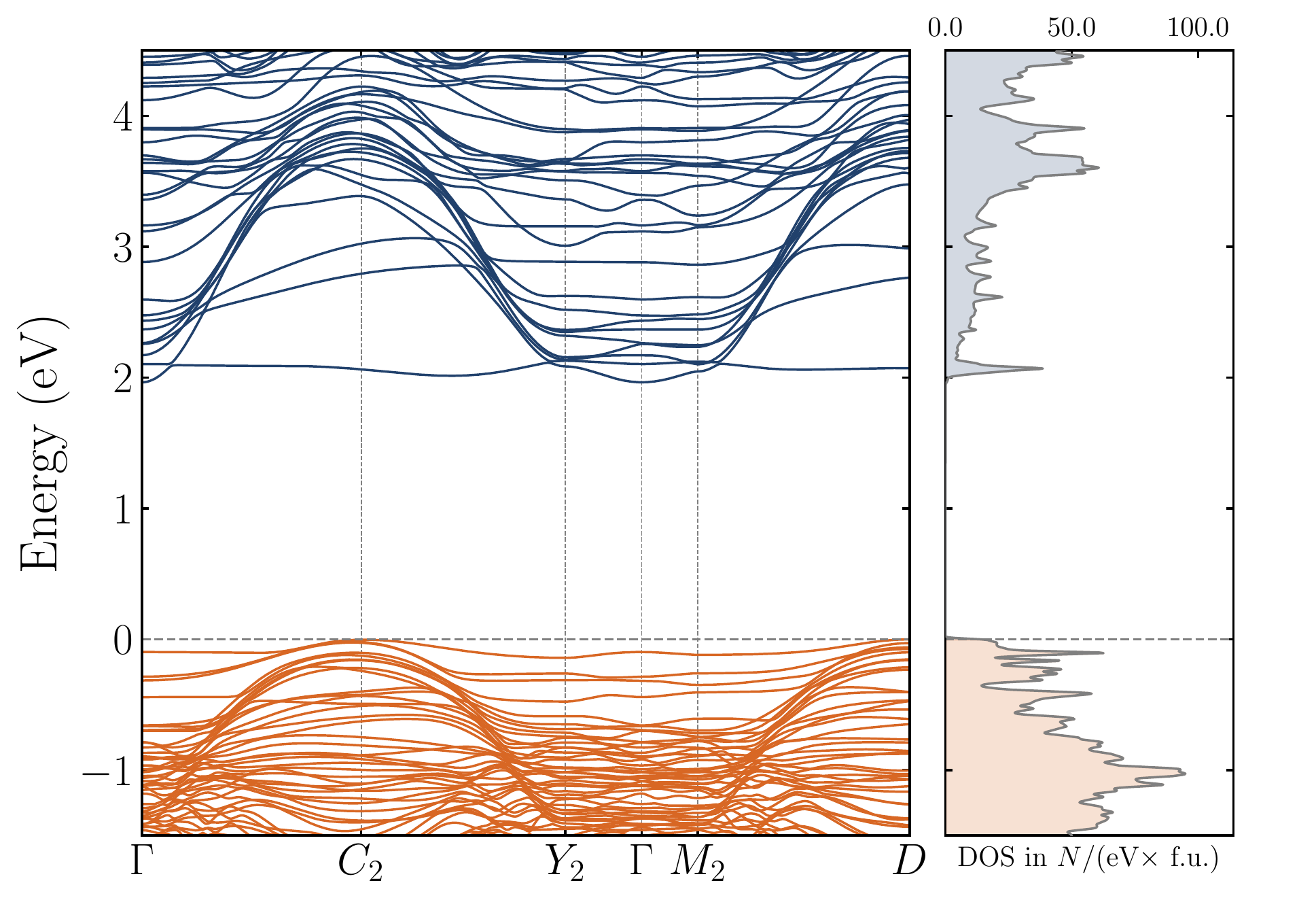}
	\caption{Bandstructure and electronic densities of states for \ce{Nb14W3O44}. Oxygen $2p$ dominated valence band is coloured in orange, while the Nb $4d$/W $5d$ conduction band is shown in blue. Both flat and dispersive conduction bands are present. The long band structure path segments involve changes in wavevector $\mathbf{k}$ along the direction reciprocal to the lattice parameter perpendicular to the block plane ($\mathbf{c}^*$ for \ce{Nb14W3O44}). The Fermi level (dashed line) sits on top of the valence band.}
	\label{fig:ElecStruc_pristine}
\end{figure}

The pristine shear phases are wide bandgap insulators (Fig.~\ref{fig:ElecStruc_pristine}, S12). The metal cations are fully oxidised and formally have a $d^0$ configuration. The valence and conduction bands (Fig.~\ref{fig:ElecStruc_pristine}) are of O 2$p$ and Nb $4d$/W $5d$ character, respectively. Lithium intercalation leads to $n$-type doping of the material, introducing electrons into the previously empty conduction band. To understand the electronic structure of the mixed-metal shear phases, it is useful to draw comparisons to the niobium suboxides \ce{Nb2O_{5-\delta}}, which also feature block-type crystal structures~\cite{cava1991a}. These compounds are formed by $n$-type doping of H-\ce{Nb2O5}, and show interesting properties: magnetism, which is rare in niobium oxides, flat bands around the Fermi energy, and an ability to host both localised and delocalised electrons~\cite{ruscher1991,cava1991a,lee2015,kocer2019}. We have previously shown that these features are fundamentally associated with the block-type crystal structure~\cite{kocer2019} and therefore also occur in \ce{Nb12WO33}, \ce{Nb14W3O44}, and \ce{Nb16W5O55} on $n$-doping. In fact, the bandstructures of the niobium-tungsten oxides show a strong similarity to those of the suboxides and H-\ce{Nb2O5}~\cite{kocer2019}, with both flat and dispersive conduction bands present (Fig.~\ref{fig:ElecStruc_pristine}, S12).

\begin{figure}[!htb]
	\centering
	\includegraphics[scale=0.27]{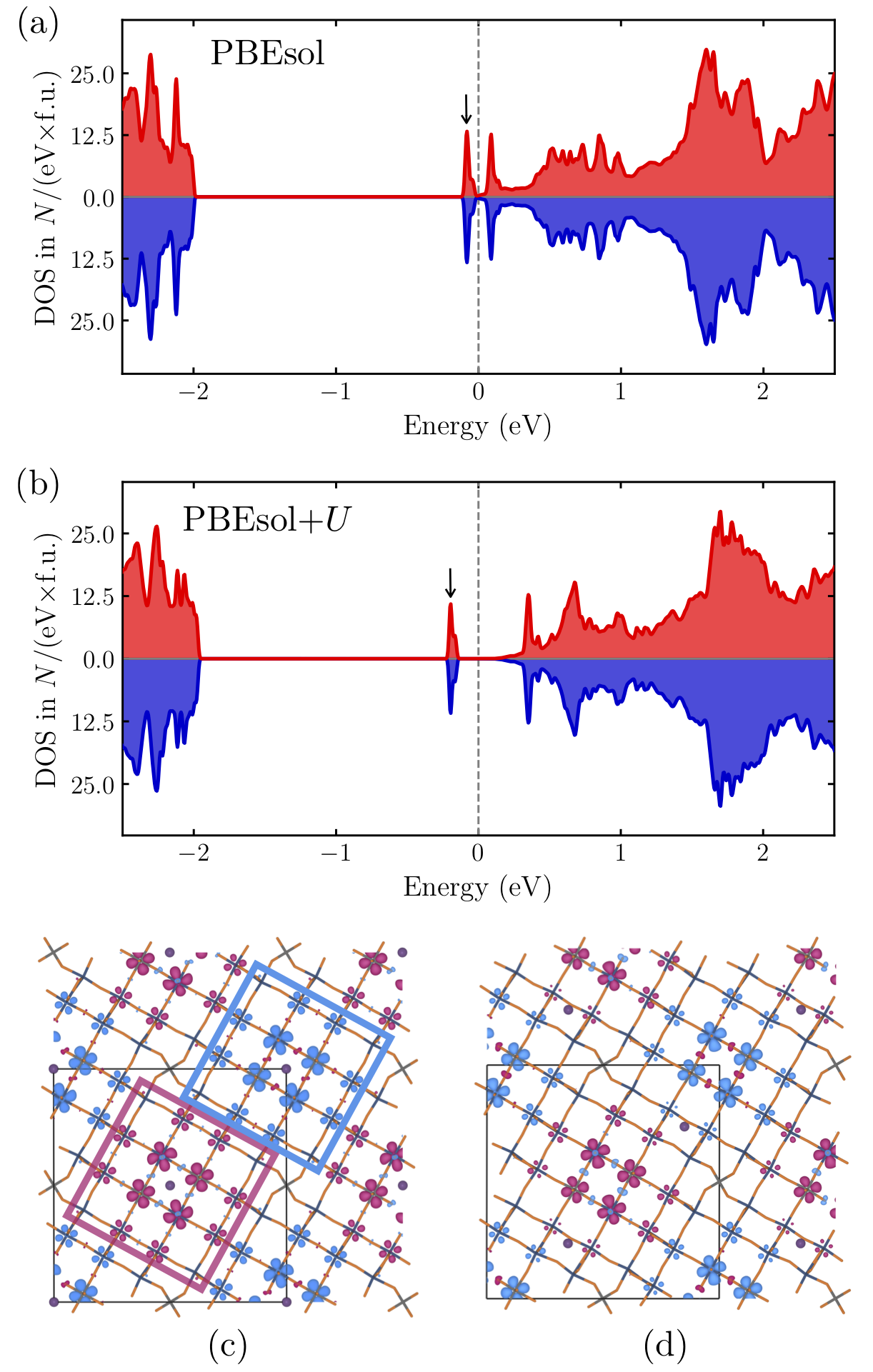}
	\caption{\ce{Li1Nb14W3O44} density of states of an antiferromagnetic spin arrangement between blocks computed with (a) PBEsol and (b) PBEsol+$U$ ($U=4$ eV). A localised state (marked by the arrow) is present in both. Spin density plots (isosurface value $0.012\ e^-/\mathrm{\AA}^3$) for structures with lithium positioned (c) in the center of the block (site 9, cf. Fig.~S8), and (d) at the edge of the block (site 1). The spin density distribution is due to the localised state shown in (a) and (b), and is independent of the lithium position.}
	\label{fig:Li1Nb14W3O44_Elec}
\end{figure}

%{\bf [\ce{Li1}]}
Insertion of a single lithium into the block of \ce{Nb14W3O44} leads to the formation of a localised electronic state (Fig.~\ref{fig:Li1Nb14W3O44_Elec}). This localised state is spread over multiple (predominantly block-central) sites and lies in the plane of the block. The localised state forms as the Fermi level is moved into the conduction band by $n$-doping, specifically by the occupation of the flat band (corresponding to the peak in the DOS, cf. Fig.~\ref{fig:ElecStruc_pristine}). A small gap is opened up between the localised state and the remainder of the conduction bands (cf. Fig.~\ref{fig:Li1Nb14W3O44_Elec}a,b; S15). Remarkably, this localisation is independent of the inclusion of a $U$ value on the Nb or W $d$-orbitals. The localisation is shown even at the GGA level, even though the gap is very small (35 meV), and increases with the introduction of a $U$ value for the metal $d$-orbitals (270 meV for $U=4$ eV). However, the spin and charge density distribution is the same. Additionally, the spin and charge distribution is also independent of whether the lithium ion is positioned in the block center or periphery (cf. Fig.~\ref{fig:Li1Nb14W3O44_Elec}c,d). This indicates that there is no strong coupling between the lithium ion and electron. A similar formation of localised electrons is also observed in \ce{Nb12WO33} and \ce{Nb16W5O55} (cf. Fig. S13). It would be interesting to determine experimentally the position of the localised dopant state relative to the bottom of the conduction band. Given that the charge associated with the localised electronic state resides predominantly on block-central sites (M1 in \ce{Nb14W3O44}, cf. Fig.~\ref{fig:NbW_cationocc}), the block interiors are reduced first upon lithium insertion into niobium-tungsten shear oxides. Since the metal positions in the block center are mostly occupied by tungsten in \ce{Nb14W3O44} and \ce{Nb16W5O55}, tungsten reduction is slightly favoured initially. In fact, this preference has been observed in \ce{Nb16W5O55} by X-ray absorption spectroscopy~\cite{griffith2018}.

\begin{figure}[!htb]
	\centering
	\includegraphics[scale=0.45]{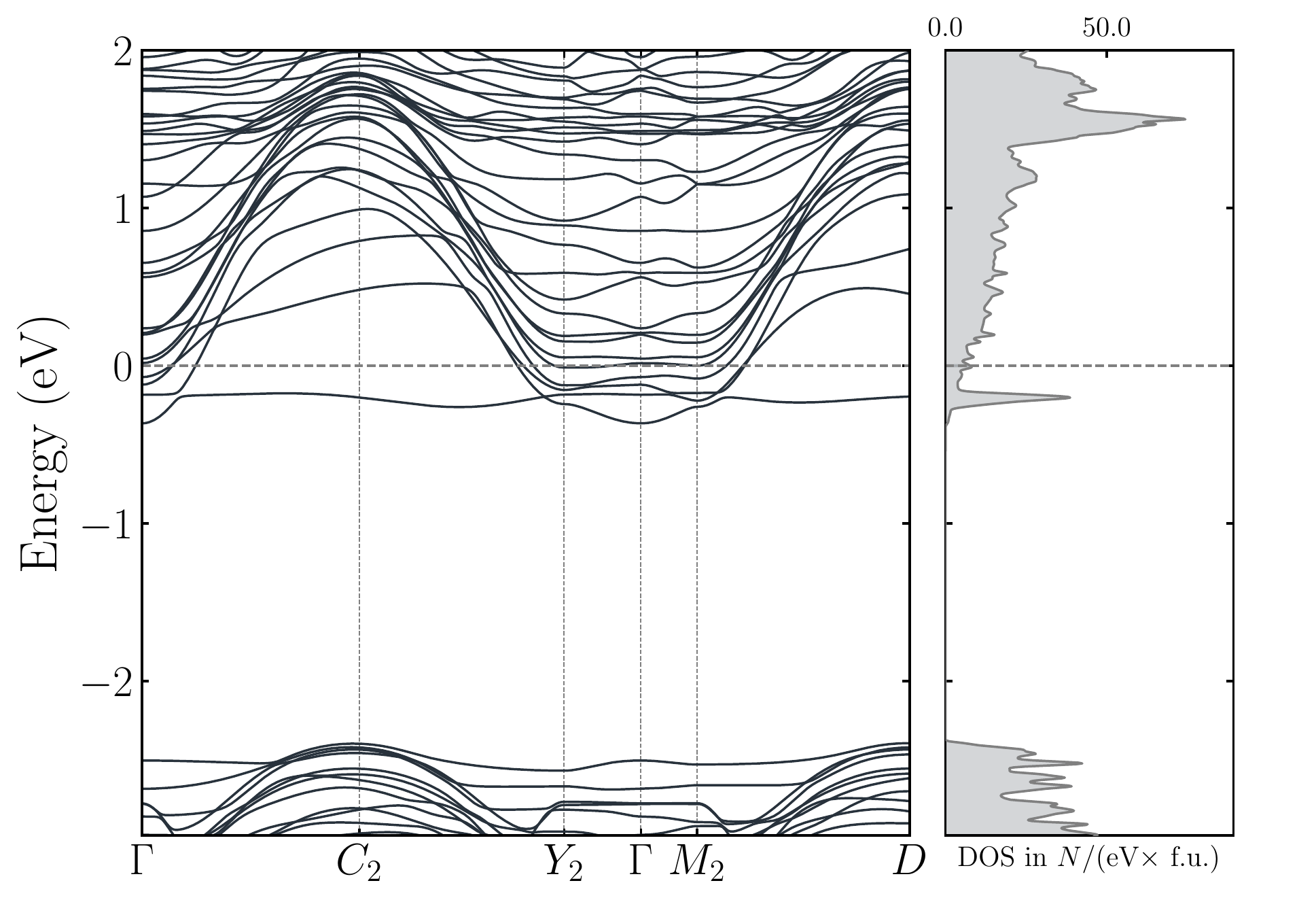}
	\caption{Bandstructure and density of states of \ce{Li3Nb14W3O44}. Relative to \ce{Nb14W3O44} (Fig.~\ref{fig:ElecStruc_pristine}), the $n$-doping by lithium insertion has moved the Fermi level (dashed line) into the conduction band.}
	\label{fig:Li3Nb14W3O44_ElecStruc}
\end{figure}

%{\bf [\ce{Li3},\ce{Li16}]}
Further $n$-doping/lithium insertion up to \ce{Li3Nb14W3O44} fully fills the flat band, but also partially fills the remaining dispersive conduction bands, resulting in metallicity (Fig.~\ref{fig:Li3Nb14W3O44_ElecStruc}). In contrast to the flat band, the dispersive conduction bands are predominantly hosted on block edge sites~\cite{kocer2019} (M2-M4 in \ce{Nb14W3O44}, cf. Fig.~\ref{fig:Li3Nb14W3O44_ElecStruc}). Reduction of the block edge sites takes place by filling these dispersive conduction bands. For even larger lithium concentrations, the structures are strongly metallic (cf. Fig. S16 for \ce{Li16Nb14W3O44}). At the GGA level, we observe no spin polarisation for either \ce{Li3Nb14W3O44} or \ce{Li16Nb14W3O44}. We do not observe the opening of a band gap by the introduction of $U$ value ($U=4$ eV) for either stoichiometry, and the compounds remain strongly metallic (Fig. S16). The same is true for fully lithiated \ce{Nb12WO33} and \ce{Nb16W5O55} (Fig. S14). Besides the slight initial preference for tungsten reduction, niobium and tungsten show similar redox activity in \ce{Nb16W5O55} (Nb$^{5+}$/Nb$^{4+}$ and W$^{6+}$/W$^{5+}$, with multielectron reduction possible beyond 1.0 Li/TM)~\cite{griffith2018}. 

%{\bf [Whats'it mean?]}
Overall, we conclude that while lithiated shear phases can show electron localisation, it is of a different type than for typical transition metal oxides. The block-structure with its orthogonal crystallographic shear planes seems to have a confinement effect such that the electron localises within the block plane, but is not confined to a single $d$-orbital on a single transition metal site. These electronic structure features are exactly the same as those observed in \ce{Nb2O_{5-\delta}}~\cite{kocer2019}. Compared to the strong localisation of small polarons in systems like \ce{Li_xTiO2}~\cite{richter2005,morgan2010} and \ce{Li_xFePO4}~\cite{maxisch2006}, the localisation in shear oxides is weaker, and easily overcome by further doping; the materials quickly become metallic on lithium insertion. The strong $d$-orbital overlap along the shear planes gives rise to large bandwidths, and in fact, the delocalised states are hosted on transition metal sites at the block periphery~\cite{kocer2019}. The preferred electron transport direction is expected to be perpendicular to the block plane, based both on experimental results on similar compounds and the calculated band dispersions~\cite{ruscher1992,kocer2019}. The good electronic conductivity suggested by these calculations is beneficial for high-rate battery performance. In addition to a good conductivity upon lithium insertion, there will be a change in the colour of the materials from white-ish to blue/black~\cite{li2018,cava1991a}. Given the facile lithiation and high-rate performance, this naturally opens up the possibility of electrochromic applications of niobium-tungsten oxides.

\newpage

\section{Discussion}

\subsection{Common Mechanistic Principles}

%{\bf [Summary]}
The three niobium-tungsten oxides \ce{Nb12WO33}, \ce{Nb14W3O44}, and \ce{Nb16W5O55} are strikingly similar in their cation ordering preferences, lithium insertion mechanisms, and electronic structure. This is expected given their close chemical and structural relationship. Regarding the lithium insertion mechanism, a set of common mechanistic principles emerge from our DFT results:

\begin{itemize}
	\item Lithium is initially inserted into fivefold coordinated sites and undistorted fourfold coordinated sites
	\item Between 0--1.5 Li/TM, the lattice evolves through three regions; the lattice parameter perpendicular to the plane of the block expands monotonically, while in the block plane, the lattice parameters expand, contract, and then expand again
	\item Distortions of the \ce{MO6} octahedra are removed over the course of lithium insertion; this symmetrisation makes previously highly distorted sites available for lithium occupation
	\item A DFT-predicted voltage profile of \ce{Nb12WO33} suggests that the lattice changes are associated with different regions of the voltage profile; during the block-plane contraction the voltage is almost constant
	\item Local and long-range structural evolution are closely linked; removal of octahedral distortions along the shear planes allows neighbouring blocks to slide closer together, causing the lattice contraction
\end{itemize}

%{\bf [Relation to Expt.]}
Experimentally, the three-region voltage profile and phase evolution is the most well-established feature of the lithiation mechanism~\cite{saritha2010,yan2017,cava1983,guo2014,yan2019,fuentes1997,griffith2017,griffith2018}. The three-stage anisotropic host-lattice response has been observed in \ce{Nb16W5O55} by Griffith et al.~\cite{griffith2018} using \textit{operando} synchrotron XRD, and correlates with the regions of the electrochemical profile. Lattice parameters of \ce{Li_xNb12WO33} phases have been reported by Cava et al.~\cite{cava1983} and Yan et al.~\cite{yan2017}. Both authors observed an anisotropic lattice change after full lithiation (\ce{Li_{10.7}Nb12WO33} and \ce{Li13Nb12WO33}, respectively), with an $a$-$c$ plane contraction and expansion along $b$. However, the lattice changes between the two studies are not consistent, with Cava et al. reporting an expansion of +8.2 \% along $b$, while Yan et al. report +3.5 \%. The study of Yan et al. was performed on nanosized material, making it not directly comparable to previous reports or DFT results.

Lattice parameters of \ce{Li_xNb14W3O44} phases have been reported by Cava et al.~\cite{cava1983}, Fuentes et al.~\cite{fuentes1997}, and Yan et al.~\cite{yan2019} While the results of Cava et al. again agree with our DFT prediction, and suggest an anisotropic evolution of the lattice parameters, the results obtained by Fuentes et al. (chemically lithiated material) and Yan et al. (nanosized material) are at variance with the DFT prediction and differ strongly from the structural evolution of the related oxides \ce{Nb12WO33} and \ce{Nb16W5O55}. We suggest that the structural evolution of \ce{Nb12WO33} and \ce{Nb14W3O44} is closer to that of \ce{Nb16W5O55} and should be re-examined. There is strong reason to believe that the similar three-region voltage profiles of \ce{Nb12WO33}, \ce{Nb14W3O44} and \ce{Nb16W5O55} are associated with a similar lattice evolution.

Regarding the local structure evolution, only results on \ce{Nb16W5O55} are available, which clearly show that the \ce{MO6} octahedra become progressively more symmetric as lithium is inserted~\cite{griffith2018}. The local structure evolution was observed through X-ray absorption spectrocopy (XAS) measurements at the Nb K-edge and W L$_{\mathrm{I}}$-edge, which show a decrease of pre-edge intensity over the course of lithium insertion. Pristine block-type crystal structures always feature strongly distorted metal-oxygen octahedra. The pre-edge arises from the dipole-forbidden $s\to d$ transition, which is absent for a metal in perfectly octahedral coordination. Removal of octahedral distortions therefore results in a decrease of intensity in this transition. Based on the DFT results, this is expected to be a universal feature of the lithium insertion mechanism of shear structures. XAS experiments on shear phase \ce{TiNb2O7} also observe such a symmetrisation in the transition metal--oxygen octahedra \cite{guo2014}, suggesting that our results are transferable to the Ti/Nb shear oxides. The reduction of $d^0$ cations prone to second-order Jahn-Teller (SOJT) distortions usually leads to a removal of the distortion (e.g. \ce{Li_xWO3} and \ce{Na_xWO3} phases~\cite{zhong1992,walkingshaw2004}). In shear oxides, the reduction can alleviate both the SOJT distortions as well as the electrostatic repulsion between cations along the shear planes, inducing symmetrisation.

\begin{figure}
    \centering
    \includegraphics[scale=0.23]{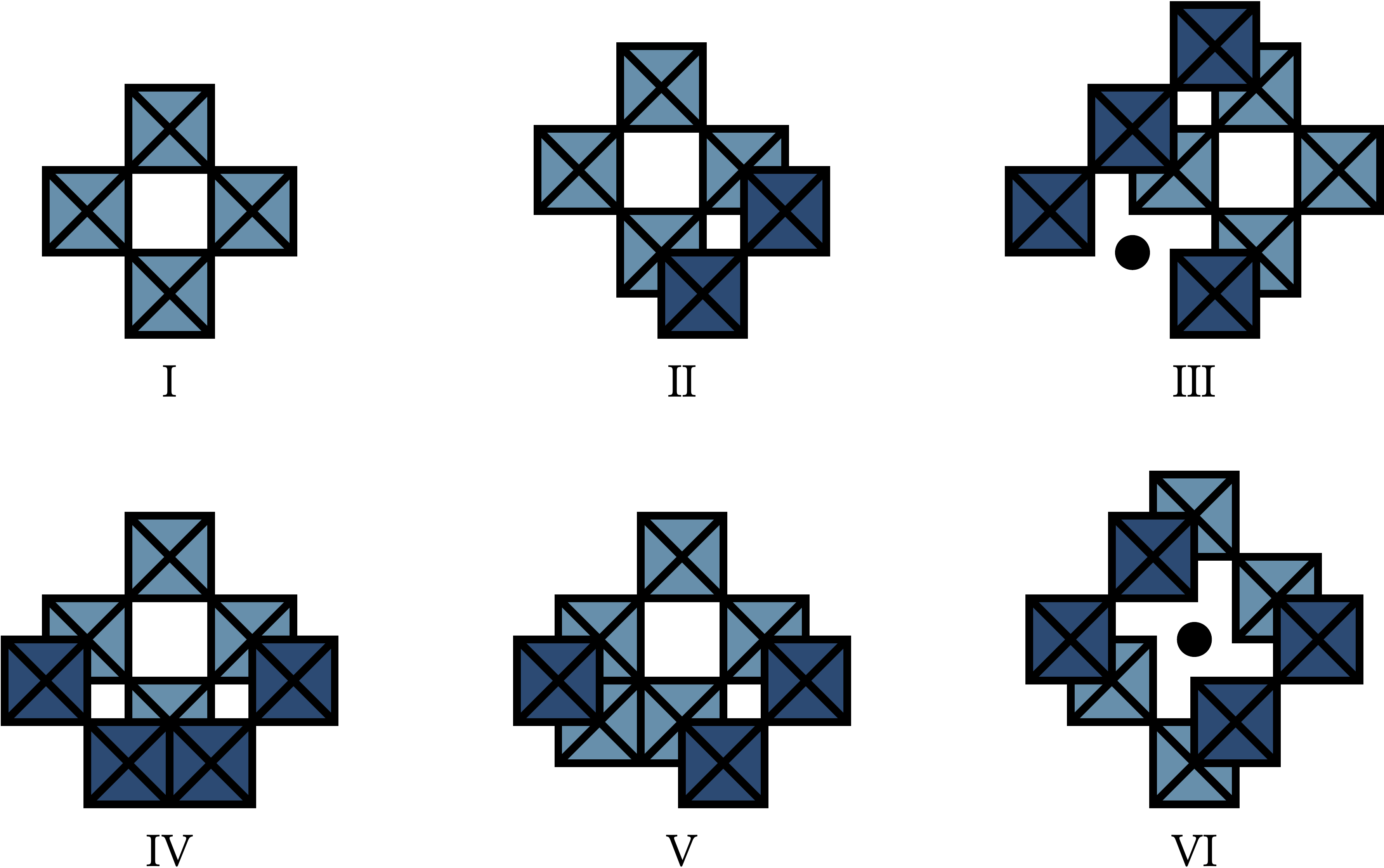}
    \caption{Cavity types found in Wadsley--Roth phases according to Cava et al. \cite{cava1983}. The tetrahedral site is denoted by a black dot.}
    \label{fig:CavityTypes}
\end{figure}

%{\bf [Cava Cavity types]}
Most previous attempts to explain the lithium insertion mechanism of block-type phases have referred to the types of cavities that are found in shear structures, which were first identified by Cava et al~\cite{cava1983}. For example, the insertion mechanism for \ce{Nb12WO33} has been proposed to proceed via insertion into type II, type III, and then type IV cavities~\cite{yan2017,saritha2010} (Figure \ref{fig:CavityTypes}). Similar mechanisms have been proposed for other block-type structures \cite{patoux2002,fuentes1997}. Our DFT calculations do not support this kind of mechanism; each cavity contains multiple lithium sites of different types (window, pocket). Instead of resorting to cavity types, it is more accurate to describe the lithium insertion mechanism by the type of site that is being filled, and what structural changes this lithium occupation causes. The cavity types are very useful, however, for the structural understanding of pristine shear oxide phases.

\subsection{Implications for Battery Performance}

We have shown that cation disorder has a significant effect on the local structure in niobium-tungsten oxide shear phases. Compared to a hypothetical ordered structure, a lithium ion within a disordered niobium tungsten oxide shear structure experiences different local environments from one unit cell to the next. The same type of lithium site (cf. Fig. S8) will be framed by different patterns of niobium and tungsten ions, with different octahedral distortions, and different local electronic structures. This randomness in the potential energy landscape of the lithium ions in a disordered structure suppresses lithium ordering and makes a larger number of sites available for occupation. While an examination of the strength of coupling between the configurations of cations and lithium ions is beyond the scope of this study, it is expected to have a beneficial effect on performance.

Given that cation disorder can be a favourable attribute to enhance electrochemical performance~\cite{griffith2018,ohno2019}, it is important to be able to control the degree of disorder. Our results suggest that tungsten energetically strongly prefers the tetrahedral site. Due to the site multiplicity and composition, \ce{Nb12WO33} can fully order with tungsten on the tetrahedral site and niobium on the block sites. However, it could be advantageous to quench from high temperatures during synthesis to lock in some degree of disorder. \ce{Nb14W3O44} and \ce{Nb16W5O55} have far more tungsten atoms than tetrahedral sites, but octahedral tungsten prefers the centre of the blocks. It would be interesting to examine the electrochemical behaviour as a function of cation disorder, controlled by the cooling rate during the synthesis of the material. Another way to increase the degree of disorder would be to introduce a third cation into the material. Within the group of $d^0$ cations titanium would be the obvious choice, since it is present in Ti/Nb crystallographic shear structures (such as \ce{TiNb2O7}). Molybdenum and zirconium would be other interesting choices.

The correlation between local and long-range structure evolution in the crystallographic shear phases directly affects the battery performance. As lithium intercalates, the total volume expansion is mitigated by the contraction within the block plane. The presence and subsequent relaxation of the octahedral distortions provides a mechanism to realise smaller volume changes in this structural family. Volume changes have an impact on long-term cycling stability; large expansion and contraction are associated with microstructural fracture, loss of particle contact within the electrode, and SEI degradation/reformation as fresh surfaces are exposed. The tempered volume changes in shear oxides thus likely contribute to their observed stability over 1000 cycles~\cite{griffith2018}, even with micrometer-dimension particles that are generally more susceptible to cracking than nanoparticles.

Many of the performance-critical properties of the niobium-tungsten oxides are intimately related to the crystal structure; the simultaneous presence of crystallographic shear planes and the \ce{ReO3}-like block interiors is key to the electrochemical performance. As previously described by other authors~\cite{cava1983,griffith2018}, the shear planes frustrate octahedral unit modes that clamp up diffusion pathways. In addition, the shear planes serve at least two other purposes: removal of local structural distortions along the shear planes buffers volume expansion, and the smaller metal-metal distances of edge-shared octahedra provide good orbital overlap and therefore enhanced electronic conductivity. The \ce{ReO3}-like block interiors, on the other hand, feature open tunnels allowing rapid lithium-ion diffusion. It seems that only when the crystal structure reaches a certain level of complexity can all of these elements be present simultaneously. The structural motifs providing each different function require structural complexity and a large unit cell size.

\section{Conclusion}

In this work, we have used an enumeration-based approach in combination with density-functional theory calculations to reveal common principles governing the cation disorder, lithium insertion mechanism, and electronic structure of the niobium-tungsten oxides \ce{Nb12WO33}, \ce{Nb14W3O44}, and \ce{Nb16W5O55}. The cross-compound transferability of our results is due to the crystallographic shear structure common to all three materials. Our results shed light on the experimentally observed three-stage lithium insertion mechanism, and reveal an important connection between the long-range and local structural changes: the removal of octahedral distortions provides a mechanism to contract the lattice in the block plane during the second stage of lithium insertion, thereby buffering the overall volume expansion. Regarding the cation disorder, we find that there is a strong preference for tungsten occupation on the tetrahedral and block-central sites of the structures. The cation disorder also has a strong influence on the local structure of the materials; different Nb/W cation arrangements produce different local octahedral distortions. Electronic structure calculations of $n$-doped/lithiated structures suggest only weak localisation of electrons upon initial lithium insertion, and the materials quickly become metallic on further lithium intercalation. Overall, our calculations suggest that the changes in local, long-range, and electronic structure on lithiation are beneficial to the battery electrode performance of the niobium-tungsten shear oxides.

Our approach of studying multiple members of one structural family has allowed us to draw compound-independent conclusions, and to use smaller model structures to represent more complex ones. The principles we have established for the niobium-tungsten shear oxides likely apply in a similar fashion to Ti/Nb oxide shear structures as well. Future computational work will focus on the extension of the mechanistic principles described here to the Ti/Nb oxide shear structures, and on modelling the diffusion process within niobium-tungsten oxide shear structures.

\acknowledgement
C.P.K. would like to thank James Darby for useful discussions. We acknowledge the use of Athena at HPC Midlands+, which was funded by the EPSRC on grant EP/P020232/1, in this research via the EPSRC RAP call of spring 2018. C.P.K. thanks the Winton Programme for the Physics of Sustainability and EPSRC for financial support. K.J.G. thanks the Winston Churchill Foundation of the United States and the Herchel Smith Foundation. K.J.G. and C.P.G. also thank the EPSRC for funding under a programme grant (EP/M009521/1). The authors declare that the data supporting the findings of this study are available within the paper and its Supporting Material files.

\suppinfo
Further details on structure enumeration and level of theory; pseudopotential specifications; tungsten site occupancies for wider temperature range; Structural evolution \ce{Nb12WO33} and \ce{Nb14W3O44} including metastable structures; Lithium sites and energies for \ce{Nb14W3O44} and \ce{Nb16W5O55}; Supplementary results on electronic structure for \ce{Nb12WO33}, \ce{Nb14W3O44}, and \ce{Nb16W5O55}; Crystallographic information files and energetics of all cation configurations of \ce{Nb14W3O44} and \ce{Nb16W5O55}; Crystallographic information files of \ce{Li_xNb12WO33} and \ce{Li_xNb14W3O44} structures.

{\scriptsize \bibliography{refs}}

\providecommand{\latin}[1]{#1}
\makeatletter
\providecommand{\doi}
  {\begingroup\let\do\@makeother\dospecials
  \catcode`\{=1 \catcode`\}=2 \doi@aux}
\providecommand{\doi@aux}[1]{\endgroup\texttt{#1}}
\makeatother
\providecommand*\mcitethebibliography{\thebibliography}
\csname @ifundefined\endcsname{endmcitethebibliography}
  {\let\endmcitethebibliography\endthebibliography}{}
\begin{mcitethebibliography}{61}
\providecommand*\natexlab[1]{#1}
\providecommand*\mciteSetBstSublistMode[1]{}
\providecommand*\mciteSetBstMaxWidthForm[2]{}
\providecommand*\mciteBstWouldAddEndPuncttrue
  {\def\EndOfBibitem{\unskip.}}
\providecommand*\mciteBstWouldAddEndPunctfalse
  {\let\EndOfBibitem\relax}
\providecommand*\mciteSetBstMidEndSepPunct[3]{}
\providecommand*\mciteSetBstSublistLabelBeginEnd[3]{}
\providecommand*\EndOfBibitem{}
\mciteSetBstSublistMode{f}
\mciteSetBstMaxWidthForm{subitem}{(\alph{mcitesubitemcount})}
\mciteSetBstSublistLabelBeginEnd
  {\mcitemaxwidthsubitemform\space}
  {\relax}
  {\relax}

\bibitem[Chen \latin{et~al.}(2013)Chen, Belharouak, Sun, and Amine]{chen2013}
Chen,~Z.; Belharouak,~I.; Sun,~Y.-K.; Amine,~K. Titanium-{{Based Anode
  Materials}} for {{Safe Lithium}}-{{Ion Batteries}}. \emph{Advanced Functional
  Materials} \textbf{2013}, \emph{23}, 959--969\relax
\mciteBstWouldAddEndPuncttrue
\mciteSetBstMidEndSepPunct{\mcitedefaultmidpunct}
{\mcitedefaultendpunct}{\mcitedefaultseppunct}\relax
\EndOfBibitem
\bibitem[Wagemaker and Mulder(2013)Wagemaker, and Mulder]{wagemaker2013}
Wagemaker,~M.; Mulder,~F.~M. Properties and {{Promises}} of {{Nanosized
  Insertion Materials}} for {{Li}}-{{Ion Batteries}}. \emph{Accounts of
  Chemical Research} \textbf{2013}, \emph{46}, 1206--1215\relax
\mciteBstWouldAddEndPuncttrue
\mciteSetBstMidEndSepPunct{\mcitedefaultmidpunct}
{\mcitedefaultendpunct}{\mcitedefaultseppunct}\relax
\EndOfBibitem
\bibitem[Griffith \latin{et~al.}(2016)Griffith, Forse, Griffin, and
  Grey]{griffith2016}
Griffith,~K.~J.; Forse,~A.~C.; Griffin,~J.~M.; Grey,~C.~P. High-{{Rate
  Intercalation}} without {{Nanostructuring}} in {{Metastable
  Nb}}{\textsubscript{2}}{{O}}{\textsubscript{5}} {{Bronze Phases}}.
  \emph{Journal of the American Chemical Society} \textbf{2016}, \emph{138},
  8888--8899\relax
\mciteBstWouldAddEndPuncttrue
\mciteSetBstMidEndSepPunct{\mcitedefaultmidpunct}
{\mcitedefaultendpunct}{\mcitedefaultseppunct}\relax
\EndOfBibitem
\bibitem[Griffith \latin{et~al.}(2018)Griffith, Wiaderek, Cibin, Marbella, and
  Grey]{griffith2018}
Griffith,~K.~J.; Wiaderek,~K.~M.; Cibin,~G.; Marbella,~L.~E.; Grey,~C.~P.
  Niobium Tungsten Oxides for High-Rate Lithium-Ion Energy Storage.
  \emph{Nature} \textbf{2018}, \emph{559}, 556--563\relax
\mciteBstWouldAddEndPuncttrue
\mciteSetBstMidEndSepPunct{\mcitedefaultmidpunct}
{\mcitedefaultendpunct}{\mcitedefaultseppunct}\relax
\EndOfBibitem
\bibitem[Griffith \latin{et~al.}(2017)Griffith, Senyshyn, and
  Grey]{griffith2017}
Griffith,~K.~J.; Senyshyn,~A.; Grey,~C.~P. Structural {{Stability}} from
  {{Crystallographic Shear}} in
  {{TiO}}{\textsubscript{2}}\textendash{{Nb}}{\textsubscript{2}}{{O}}{\textsubscript{5}}
  {{Phases}}: {{Cation Ordering}} and {{Lithiation Behavior}} of
  {{TiNb}}{\textsubscript{24}}{{O}}{\textsubscript{62}}. \emph{Inorganic
  Chemistry} \textbf{2017}, \emph{56}, 4002--4010\relax
\mciteBstWouldAddEndPuncttrue
\mciteSetBstMidEndSepPunct{\mcitedefaultmidpunct}
{\mcitedefaultendpunct}{\mcitedefaultseppunct}\relax
\EndOfBibitem
\bibitem[Kato and Tamura(1975)Kato, and Tamura]{kato1975}
Kato,~K.; Tamura,~S. {Die Kristallstruktur von
  T-Nb\textsubscript{2}O\textsubscript{5}}. \emph{Acta Crystallographica
  Section B: Structural Crystallography and Crystal Chemistry} \textbf{1975},
  \emph{31}, 673--677\relax
\mciteBstWouldAddEndPuncttrue
\mciteSetBstMidEndSepPunct{\mcitedefaultmidpunct}
{\mcitedefaultendpunct}{\mcitedefaultseppunct}\relax
\EndOfBibitem
\bibitem[Roth and Wadsley(1965)Roth, and Wadsley]{roth1965a}
Roth,~R.~S.; Wadsley,~A.~D. Multiple Phase Formation in the Binary System
  {{Nb}}{\textsubscript{2}}{{O}}{\textsubscript{5}}\textendash{{WO}}{\textsubscript{3}}.
  {{I}}. {{Preparation}} and Identification of Phases. \emph{Acta
  Crystallographica} \textbf{1965}, \emph{19}, 26--32\relax
\mciteBstWouldAddEndPuncttrue
\mciteSetBstMidEndSepPunct{\mcitedefaultmidpunct}
{\mcitedefaultendpunct}{\mcitedefaultseppunct}\relax
\EndOfBibitem
\bibitem[Wadsley(1961)]{wadsley1961}
Wadsley,~A.~D. Mixed Oxides of Titanium and Niobium. {{I}}. \emph{Acta
  Crystallographica} \textbf{1961}, \emph{14}, 660--664\relax
\mciteBstWouldAddEndPuncttrue
\mciteSetBstMidEndSepPunct{\mcitedefaultmidpunct}
{\mcitedefaultendpunct}{\mcitedefaultseppunct}\relax
\EndOfBibitem
\bibitem[Kato(1976)]{kato1976}
Kato,~K. Structure Refinement of
  {{H}}-{{Nb}}{\textsubscript{2}}{{O}}{\textsubscript{5}}. \emph{Acta
  Crystallographica Section B: Structural Crystallography and Crystal
  Chemistry} \textbf{1976}, \emph{32}, 764--767\relax
\mciteBstWouldAddEndPuncttrue
\mciteSetBstMidEndSepPunct{\mcitedefaultmidpunct}
{\mcitedefaultendpunct}{\mcitedefaultseppunct}\relax
\EndOfBibitem
\bibitem[Cava \latin{et~al.}(1991)Cava, Batlogg, Krajewski, Poulsen, Gammel,
  Peck, and Rupp]{cava1991a}
Cava,~R.~J.; Batlogg,~B.; Krajewski,~J.~J.; Poulsen,~H.~F.; Gammel,~P.;
  Peck,~W.~F.; Rupp,~L.~W. Electrical and Magnetic Properties of
  {{Nb}}{\textsubscript{2}}{{O}}{\textsubscript{5-{$\delta$}}} Crystallographic
  Shear Structures. \emph{Physical Review B} \textbf{1991}, \emph{44},
  6973--6981\relax
\mciteBstWouldAddEndPuncttrue
\mciteSetBstMidEndSepPunct{\mcitedefaultmidpunct}
{\mcitedefaultendpunct}{\mcitedefaultseppunct}\relax
\EndOfBibitem
\bibitem[Kunz and Brown(1995)Kunz, and Brown]{kunz1995}
Kunz,~M.; Brown,~I.~D. Out-of-{{Center Distortions}} around {{Octahedrally
  Coordinated}} {\emph{D}}{\textsuperscript{0}} {{Transition Metals}}.
  \emph{Journal of Solid State Chemistry} \textbf{1995}, \emph{115},
  395--406\relax
\mciteBstWouldAddEndPuncttrue
\mciteSetBstMidEndSepPunct{\mcitedefaultmidpunct}
{\mcitedefaultendpunct}{\mcitedefaultseppunct}\relax
\EndOfBibitem
\bibitem[Bersuker(2006)]{bersuker2006}
Bersuker,~I. \emph{The {{Jahn}}\textendash{{Teller Effect}}}; {Cambridge
  University Press}: {Cambridge}, 2006\relax
\mciteBstWouldAddEndPuncttrue
\mciteSetBstMidEndSepPunct{\mcitedefaultmidpunct}
{\mcitedefaultendpunct}{\mcitedefaultseppunct}\relax
\EndOfBibitem
\bibitem[Cava \latin{et~al.}(1983)Cava, Murphy, and Zahurak]{cava1983}
Cava,~R.~J.; Murphy,~D.~W.; Zahurak,~S.~M. Lithium {{Insertion}} in
  {{Wadsley}}-{{Roth Phases Based}} on {{Niobium Oxide}}. \emph{Journal of The
  Electrochemical Society} \textbf{1983}, \emph{130}, 2345--2351\relax
\mciteBstWouldAddEndPuncttrue
\mciteSetBstMidEndSepPunct{\mcitedefaultmidpunct}
{\mcitedefaultendpunct}{\mcitedefaultseppunct}\relax
\EndOfBibitem
\bibitem[Cava \latin{et~al.}(1981)Cava, Santoro, Murphy, Zahurak, and
  Roth]{cava1981}
Cava,~R.~J.; Santoro,~A.; Murphy,~D.~W.; Zahurak,~S.; Roth,~R.~S. Structural
  Aspects of Lithium Insertion in Oxides:
  {{Li}}{\textsubscript{x}}{{ReO}}{\textsubscript{3}} and
  {{Li}}{\textsubscript{2}}{{FeV}}{\textsubscript{3}}{{O}}{\textsubscript{8}}.
  \emph{Solid State Ionics} \textbf{1981}, \emph{5}, 323--326\relax
\mciteBstWouldAddEndPuncttrue
\mciteSetBstMidEndSepPunct{\mcitedefaultmidpunct}
{\mcitedefaultendpunct}{\mcitedefaultseppunct}\relax
\EndOfBibitem
\bibitem[Lu \latin{et~al.}(2011)Lu, Jian, Fang, Gu, Hu, Chen, Wang, and
  Chen]{lu2011}
Lu,~X.; Jian,~Z.; Fang,~Z.; Gu,~L.; Hu,~Y.-S.; Chen,~W.; Wang,~Z.; Chen,~L.
  Atomic-Scale Investigation on Lithium Storage Mechanism in
  {{TiNb}}{\textsubscript{2}}{{O}}{\textsubscript{7}},. \emph{Energy \&
  Environmental Science} \textbf{2011}, \emph{4}, 2638--2644\relax
\mciteBstWouldAddEndPuncttrue
\mciteSetBstMidEndSepPunct{\mcitedefaultmidpunct}
{\mcitedefaultendpunct}{\mcitedefaultseppunct}\relax
\EndOfBibitem
\bibitem[Guo \latin{et~al.}(2014)Guo, Yu, Sun, Chi, Qiao, Liu, Hu, Yang,
  Goodenough, and Dai]{guo2014}
Guo,~B.; Yu,~X.; Sun,~X.-G.; Chi,~M.; Qiao,~Z.-A.; Liu,~J.; Hu,~Y.-S.;
  Yang,~X.-Q.; Goodenough,~J.~B.; Dai,~S. A Long-Life Lithium-Ion Battery with
  a Highly Porous {{TiNb}}{\textsubscript{2}}{{O}}{\textsubscript{7}} Anode for
  Large-Scale Electrical Energy Storage. \emph{Energy Environ. Sci.}
  \textbf{2014}, \emph{7}, 2220--2226\relax
\mciteBstWouldAddEndPuncttrue
\mciteSetBstMidEndSepPunct{\mcitedefaultmidpunct}
{\mcitedefaultendpunct}{\mcitedefaultseppunct}\relax
\EndOfBibitem
\bibitem[Cheng \latin{et~al.}(2014)Cheng, Liang, Zhu, Si, Guo, and
  Qian]{cheng2014}
Cheng,~Q.; Liang,~J.; Zhu,~Y.; Si,~L.; Guo,~C.; Qian,~Y. Bulk
  {{Ti}}{\textsubscript{2}}{{Nb}}{\textsubscript{10}}{{O}}{\textsubscript{29}}
  as Long-Life and High-Power {{Li}}-Ion Battery Anodes. \emph{J. Mater. Chem.
  A} \textbf{2014}, \emph{2}, 17258--17262\relax
\mciteBstWouldAddEndPuncttrue
\mciteSetBstMidEndSepPunct{\mcitedefaultmidpunct}
{\mcitedefaultendpunct}{\mcitedefaultseppunct}\relax
\EndOfBibitem
\bibitem[Wu \latin{et~al.}(2012)Wu, Miao, Han, Hu, Chen, Lee, Kim, and
  Chen]{wu2012}
Wu,~X.; Miao,~J.; Han,~W.; Hu,~Y.-S.; Chen,~D.; Lee,~J.-S.; Kim,~J.; Chen,~L.
  Investigation on
  {{Ti}}{\textsubscript{2}}{{Nb}}{\textsubscript{10}}{{O}}{\textsubscript{29}}
  Anode Material for Lithium-Ion Batteries. \emph{Electrochemistry
  Communications} \textbf{2012}, \emph{25}, 39--42\relax
\mciteBstWouldAddEndPuncttrue
\mciteSetBstMidEndSepPunct{\mcitedefaultmidpunct}
{\mcitedefaultendpunct}{\mcitedefaultseppunct}\relax
\EndOfBibitem
\bibitem[Saritha \latin{et~al.}(2010)Saritha, Pralong, Varadaraju, and
  Raveau]{saritha2010}
Saritha,~D.; Pralong,~V.; Varadaraju,~U.~V.; Raveau,~B. Electrochemical {{Li}}
  Insertion Studies on
  {{WNb}}{\textsubscript{12}}{{O}}{\textsubscript{33}}\textemdash{{A}} Shear
  {{ReO}}{\textsubscript{3}} Type Structure. \emph{Journal of Solid State
  Chemistry} \textbf{2010}, \emph{183}, 988--993\relax
\mciteBstWouldAddEndPuncttrue
\mciteSetBstMidEndSepPunct{\mcitedefaultmidpunct}
{\mcitedefaultendpunct}{\mcitedefaultseppunct}\relax
\EndOfBibitem
\bibitem[Yan \latin{et~al.}(2017)Yan, Lan, Yu, Qian, Cheng, Long, Zhang, Shui,
  and Shu]{yan2017}
Yan,~L.; Lan,~H.; Yu,~H.; Qian,~S.; Cheng,~X.; Long,~N.; Zhang,~R.; Shui,~M.;
  Shu,~J. Electrospun {{WNb}}{\textsubscript{12}}{{O}}{\textsubscript{33}}
  Nanowires: Superior Lithium Storage Capability and Their Working Mechanism.
  \emph{Journal of Materials Chemistry A} \textbf{2017}, \emph{5},
  8972--8980\relax
\mciteBstWouldAddEndPuncttrue
\mciteSetBstMidEndSepPunct{\mcitedefaultmidpunct}
{\mcitedefaultendpunct}{\mcitedefaultseppunct}\relax
\EndOfBibitem
\bibitem[Fuentes \latin{et~al.}(1997)Fuentes, Garza, {de la Cruz}, and
  {Torres-Mart{\'i}nez}]{fuentes1997}
Fuentes,~A.~F.; Garza,~E.~B.; {de la Cruz},~A.~M.; {Torres-Mart{\'i}nez},~L.~M.
  Lithium and Sodium Insertion in
  {{W}}{\textsubscript{3}}{{Nb}}{\textsubscript{14}}{{O}}{\textsubscript{44}},
  a Block Structure Type Phase. \emph{Solid state ionics} \textbf{1997},
  \emph{93}, 245--253\relax
\mciteBstWouldAddEndPuncttrue
\mciteSetBstMidEndSepPunct{\mcitedefaultmidpunct}
{\mcitedefaultendpunct}{\mcitedefaultseppunct}\relax
\EndOfBibitem
\bibitem[Yan \latin{et~al.}(2019)Yan, Shu, Li, Cheng, Zhu, Yu, Zhang, Zheng,
  Xie, and Guo]{yan2019}
Yan,~L.; Shu,~J.; Li,~C.; Cheng,~X.; Zhu,~H.; Yu,~H.; Zhang,~C.; Zheng,~Y.;
  Xie,~Y.; Guo,~Z.
  W{\textsubscript{3}}{{Nb}}{\textsubscript{14}}{{O}}{\textsubscript{44}}
  Nanowires: {{Ultrastable}} Lithium Storage Anode Materials for Advanced
  Rechargeable Batteries. \emph{Energy Storage Materials} \textbf{2019},
  \emph{16}, 535--544\relax
\mciteBstWouldAddEndPuncttrue
\mciteSetBstMidEndSepPunct{\mcitedefaultmidpunct}
{\mcitedefaultendpunct}{\mcitedefaultseppunct}\relax
\EndOfBibitem
\bibitem[Li \latin{et~al.}(2018)Li, Qin, Liu, Yang, Lin, Xia, Lin, Chen, and
  Li]{li2018}
Li,~R.; Qin,~Y.; Liu,~X.; Yang,~L.; Lin,~C.; Xia,~R.; Lin,~S.; Chen,~Y.; Li,~J.
  Conductive {{Nb}}{\textsubscript{25}}{{O}}{\textsubscript{62}} and
  {{Nb}}{\textsubscript{12}}{{O}}{\textsubscript{29}} Anode Materials for Use
  in High-Performance Lithium-Ion Storage. \emph{Electrochimica Acta}
  \textbf{2018}, \emph{266}, 202--211\relax
\mciteBstWouldAddEndPuncttrue
\mciteSetBstMidEndSepPunct{\mcitedefaultmidpunct}
{\mcitedefaultendpunct}{\mcitedefaultseppunct}\relax
\EndOfBibitem
\bibitem[Li \latin{et~al.}(2011)Li, Sun, and Goodenough]{li2011}
Li,~Y.; Sun,~C.; Goodenough,~J.~B. Electrochemical {{Lithium Intercalation}} in
  {{Monoclinic Nb}}{\textsubscript{12}}{{O}}{\textsubscript{29}}.
  \emph{Chemistry of Materials} \textbf{2011}, \emph{23}, 2292--2294\relax
\mciteBstWouldAddEndPuncttrue
\mciteSetBstMidEndSepPunct{\mcitedefaultmidpunct}
{\mcitedefaultendpunct}{\mcitedefaultseppunct}\relax
\EndOfBibitem
\bibitem[Patoux \latin{et~al.}(2002)Patoux, Dolle, Rousse, and
  Masquelier]{patoux2002}
Patoux,~S.; Dolle,~M.; Rousse,~G.; Masquelier,~C. A {{Reversible Lithium
  Intercalation Process}} in an {{ReO}}{\textsubscript{3}} \- {{Type Structure
  PNb}}{\textsubscript{9}}{{O}}{\textsubscript{25}}. \emph{Journal of The
  Electrochemical Society} \textbf{2002}, \emph{149}, A391--A400\relax
\mciteBstWouldAddEndPuncttrue
\mciteSetBstMidEndSepPunct{\mcitedefaultmidpunct}
{\mcitedefaultendpunct}{\mcitedefaultseppunct}\relax
\EndOfBibitem
\bibitem[Cheetham and Dreele(1973)Cheetham, and Dreele]{cheetham1973}
Cheetham,~A.~K.; Dreele,~R. B.~V. Cation {{Distributions}} in {{Niobium Oxide
  Block Structures}}. \emph{Nature Physical Science} \textbf{1973}, \emph{244},
  139--140\relax
\mciteBstWouldAddEndPuncttrue
\mciteSetBstMidEndSepPunct{\mcitedefaultmidpunct}
{\mcitedefaultendpunct}{\mcitedefaultseppunct}\relax
\EndOfBibitem
\bibitem[Ko{\c c}er \latin{et~al.}(2019)Ko{\c c}er, Griffith, Grey, and
  Morris]{kocer2019}
Ko{\c c}er,~C.~P.; Griffith,~K.~J.; Grey,~C.~P.; Morris,~A.~J. First-Principles
  Study of Localized and Delocalized Electronic States in Crystallographic
  Shear Phases of Niobium Oxide. \emph{Physical Review B} \textbf{2019},
  \emph{99}, 075151\relax
\mciteBstWouldAddEndPuncttrue
\mciteSetBstMidEndSepPunct{\mcitedefaultmidpunct}
{\mcitedefaultendpunct}{\mcitedefaultseppunct}\relax
\EndOfBibitem
\bibitem[{Grau-Crespo} \latin{et~al.}(2007){Grau-Crespo}, Hamad, Catlow, and
  de~Leeuw]{grau-crespo2007}
{Grau-Crespo},~R.; Hamad,~S.; Catlow,~C. R.~A.; de~Leeuw,~N.~H.
  Symmetry-Adapted Configurational Modelling of Fractional Site Occupancy in
  Solids. \emph{Journal of Physics: Condensed Matter} \textbf{2007}, \emph{19},
  256201\relax
\mciteBstWouldAddEndPuncttrue
\mciteSetBstMidEndSepPunct{\mcitedefaultmidpunct}
{\mcitedefaultendpunct}{\mcitedefaultseppunct}\relax
\EndOfBibitem
\bibitem[Clark \latin{et~al.}(2005)Clark, Segall, Pickard, Hasnip, Probert,
  Refson, and Payne]{clark2005}
Clark,~S.~J.; Segall,~M.~D.; Pickard,~C.~J.; Hasnip,~P.~J.; Probert,~M. I.~J.;
  Refson,~K.; Payne,~M.~C. First Principles Methods Using {{CASTEP}}.
  \emph{Zeitschrift f{\"u}r Kristallographie - Crystalline Materials}
  \textbf{2005}, \emph{220}, 567--570\relax
\mciteBstWouldAddEndPuncttrue
\mciteSetBstMidEndSepPunct{\mcitedefaultmidpunct}
{\mcitedefaultendpunct}{\mcitedefaultseppunct}\relax
\EndOfBibitem
\bibitem[Perdew \latin{et~al.}(2008)Perdew, Ruzsinszky, Csonka, Vydrov,
  Scuseria, Constantin, Zhou, and Burke]{perdew2008}
Perdew,~J.~P.; Ruzsinszky,~A.; Csonka,~G.~I.; Vydrov,~O.~A.; Scuseria,~G.~E.;
  Constantin,~L.~A.; Zhou,~X.; Burke,~K. Restoring the {{Density}}-{{Gradient
  Expansion}} for {{Exchange}} in {{Solids}} and {{Surfaces}}. \emph{Physical
  Review Letters} \textbf{2008}, \emph{100}, 136406\relax
\mciteBstWouldAddEndPuncttrue
\mciteSetBstMidEndSepPunct{\mcitedefaultmidpunct}
{\mcitedefaultendpunct}{\mcitedefaultseppunct}\relax
\EndOfBibitem
\bibitem[Vanderbilt(1990)]{vanderbilt1990}
Vanderbilt,~D. Soft Self-Consistent Pseudopotentials in a Generalized
  Eigenvalue Formalism. \emph{Physical Review B} \textbf{1990}, \emph{41},
  7892--7895\relax
\mciteBstWouldAddEndPuncttrue
\mciteSetBstMidEndSepPunct{\mcitedefaultmidpunct}
{\mcitedefaultendpunct}{\mcitedefaultseppunct}\relax
\EndOfBibitem
\bibitem[Monkhorst and Pack(1976)Monkhorst, and Pack]{monkhorst1976}
Monkhorst,~H.~J.; Pack,~J.~D. Special Points for {{Brillouin}}-Zone
  Integrations. \emph{Physical Review B} \textbf{1976}, \emph{13},
  5188--5192\relax
\mciteBstWouldAddEndPuncttrue
\mciteSetBstMidEndSepPunct{\mcitedefaultmidpunct}
{\mcitedefaultendpunct}{\mcitedefaultseppunct}\relax
\EndOfBibitem
\bibitem[Dudarev \latin{et~al.}(1998)Dudarev, Botton, Savrasov, Humphreys, and
  Sutton]{dudarev1998}
Dudarev,~S.~L.; Botton,~G.~A.; Savrasov,~S.~Y.; Humphreys,~C.~J.; Sutton,~A.~P.
  Electron-Energy-Loss Spectra and the Structural Stability of Nickel Oxide:
  {{An LSDA}}+{{U}} Study. \emph{Physical Review B} \textbf{1998}, \emph{57},
  1505--1509\relax
\mciteBstWouldAddEndPuncttrue
\mciteSetBstMidEndSepPunct{\mcitedefaultmidpunct}
{\mcitedefaultendpunct}{\mcitedefaultseppunct}\relax
\EndOfBibitem
\bibitem[Aydinol \latin{et~al.}(1997)Aydinol, Kohan, Ceder, Cho, and
  Joannopoulos]{aydinol1997}
Aydinol,~M.~K.; Kohan,~A.~F.; Ceder,~G.; Cho,~K.; Joannopoulos,~J. Ab Initio
  Study of Lithium Intercalation in Metal Oxides and Metal Dichalcogenides.
  \emph{Physical Review B} \textbf{1997}, \emph{56}, 1354--1365\relax
\mciteBstWouldAddEndPuncttrue
\mciteSetBstMidEndSepPunct{\mcitedefaultmidpunct}
{\mcitedefaultendpunct}{\mcitedefaultseppunct}\relax
\EndOfBibitem
\bibitem[Hinuma \latin{et~al.}(2017)Hinuma, Pizzi, Kumagai, Oba, and
  Tanaka]{hinuma2017}
Hinuma,~Y.; Pizzi,~G.; Kumagai,~Y.; Oba,~F.; Tanaka,~I. Band Structure Diagram
  Paths Based on Crystallography. \emph{Computational Materials Science}
  \textbf{2017}, \emph{128}, 140--184\relax
\mciteBstWouldAddEndPuncttrue
\mciteSetBstMidEndSepPunct{\mcitedefaultmidpunct}
{\mcitedefaultendpunct}{\mcitedefaultseppunct}\relax
\EndOfBibitem
\bibitem[Togo and Tanaka(2018)Togo, and Tanaka]{togo2018}
Togo,~A.; Tanaka,~I. Spglib: A Software Library for Crystal Symmetry Search.
  \emph{arXiv:1808.01590 [cond-mat]} \textbf{2018}, \relax
\mciteBstWouldAddEndPunctfalse
\mciteSetBstMidEndSepPunct{\mcitedefaultmidpunct}
{}{\mcitedefaultseppunct}\relax
\EndOfBibitem
\bibitem[Morris \latin{et~al.}(2014)Morris, Nicholls, Pickard, and
  Yates]{morris2014a}
Morris,~A.~J.; Nicholls,~R.~J.; Pickard,~C.~J.; Yates,~J.~R. {{OptaDOS}}: {{A}}
  Tool for Obtaining Density of States, Core-Level and Optical Spectra from
  Electronic Structure Codes. \emph{Computer Physics Communications}
  \textbf{2014}, \emph{185}, 1477--1485\relax
\mciteBstWouldAddEndPuncttrue
\mciteSetBstMidEndSepPunct{\mcitedefaultmidpunct}
{\mcitedefaultendpunct}{\mcitedefaultseppunct}\relax
\EndOfBibitem
\bibitem[Pickard and Payne(1999)Pickard, and Payne]{pickard1999}
Pickard,~C.~J.; Payne,~M.~C. Extrapolative Approaches to {{Brillouin}}-Zone
  Integration. \emph{Physical Review B} \textbf{1999}, \emph{59},
  4685--4693\relax
\mciteBstWouldAddEndPuncttrue
\mciteSetBstMidEndSepPunct{\mcitedefaultmidpunct}
{\mcitedefaultendpunct}{\mcitedefaultseppunct}\relax
\EndOfBibitem
\bibitem[Pickard and Payne(2000)Pickard, and Payne]{pickard2000}
Pickard,~C.~J.; Payne,~M.~C. Second-Order k {$\cdot$} p Perturbation Theory
  with {{Vanderbilt}} Pseudopotentials and Plane Waves. \emph{Physical Review
  B} \textbf{2000}, \emph{62}, 4383--4388\relax
\mciteBstWouldAddEndPuncttrue
\mciteSetBstMidEndSepPunct{\mcitedefaultmidpunct}
{\mcitedefaultendpunct}{\mcitedefaultseppunct}\relax
\EndOfBibitem
\bibitem[Rutter(2018)]{rutter2018}
Rutter,~M.~J. C2x: {{A}} Tool for Visualisation and Input Preparation for
  {{Castep}} and Other Electronic Structure Codes. \emph{Computer Physics
  Communications} \textbf{2018}, \emph{225}, 174--179\relax
\mciteBstWouldAddEndPuncttrue
\mciteSetBstMidEndSepPunct{\mcitedefaultmidpunct}
{\mcitedefaultendpunct}{\mcitedefaultseppunct}\relax
\EndOfBibitem
\bibitem[Momma and Izumi(2011)Momma, and Izumi]{momma2011}
Momma,~K.; Izumi,~F. {{VESTA}} 3 for Three-Dimensional Visualization of
  Crystal, Volumetric and Morphology Data. \emph{Journal of Applied
  Crystallography} \textbf{2011}, \emph{44}, 1272--1276\relax
\mciteBstWouldAddEndPuncttrue
\mciteSetBstMidEndSepPunct{\mcitedefaultmidpunct}
{\mcitedefaultendpunct}{\mcitedefaultseppunct}\relax
\EndOfBibitem
\bibitem[Evans()]{evans}
Evans,~M. Ml-Evs / Matador. https://bitbucket.org/ml-evs/matador\relax
\mciteBstWouldAddEndPuncttrue
\mciteSetBstMidEndSepPunct{\mcitedefaultmidpunct}
{\mcitedefaultendpunct}{\mcitedefaultseppunct}\relax
\EndOfBibitem
\bibitem[Cheetham and Allen(1983)Cheetham, and Allen]{cheetham1983}
Cheetham,~A.~K.; Allen,~N.~C. Cation Distribution in the Complex Oxide,
  {{W}}{\textsubscript{3}}{{Nb}}{\textsubscript{14}}{{O}}{\textsubscript{44}};
  a Time-of-Flight Neutron Diffraction Study. \emph{Journal of the Chemical
  Society, Chemical Communications} \textbf{1983}, \emph{0}, 1370--1372\relax
\mciteBstWouldAddEndPuncttrue
\mciteSetBstMidEndSepPunct{\mcitedefaultmidpunct}
{\mcitedefaultendpunct}{\mcitedefaultseppunct}\relax
\EndOfBibitem
\bibitem[Roth and Wadsley(1965)Roth, and Wadsley]{roth1965b}
Roth,~R.~S.; Wadsley,~A.~D. Multiple Phase Formation in the Binary System
  {{Nb}}{\textsubscript{2}}{{O}}{\textsubscript{5}}-{{WO}}{\textsubscript{3}}.
  {{II}}. {{The}} Structure of the Monoclinic Phases
  {{WNb}}{\textsubscript{12}}{{O}}{\textsubscript{33}} and
  {{W}}{\textsubscript{5}}{{Nb}}{\textsubscript{16}}{{O}}{\textsubscript{55}}.
  \emph{Acta Crystallographica} \textbf{1965}, \emph{19}, 32--38\relax
\mciteBstWouldAddEndPuncttrue
\mciteSetBstMidEndSepPunct{\mcitedefaultmidpunct}
{\mcitedefaultendpunct}{\mcitedefaultseppunct}\relax
\EndOfBibitem
\bibitem[Sears(1992)]{sears1992}
Sears,~V.~F. Neutron Scattering Lengths and Cross Sections. \emph{Neutron News}
  \textbf{1992}, \emph{3}, 26--37\relax
\mciteBstWouldAddEndPuncttrue
\mciteSetBstMidEndSepPunct{\mcitedefaultmidpunct}
{\mcitedefaultendpunct}{\mcitedefaultseppunct}\relax
\EndOfBibitem
\bibitem[Robinson \latin{et~al.}(1971)Robinson, Gibbs, and Ribbe]{robinson1971}
Robinson,~K.; Gibbs,~G.~V.; Ribbe,~P.~H. Quadratic {{Elongation}}: {{A
  Quantitative Measure}} of {{Distortion}} in {{Coordination Polyhedra}}.
  \emph{Science} \textbf{1971}, \emph{172}, 567--570\relax
\mciteBstWouldAddEndPuncttrue
\mciteSetBstMidEndSepPunct{\mcitedefaultmidpunct}
{\mcitedefaultendpunct}{\mcitedefaultseppunct}\relax
\EndOfBibitem
\bibitem[Ok \latin{et~al.}(2006)Ok, Halasyamani, Casanova, Llunell, Alemany,
  and Alvarez]{ok2006}
Ok,~K.~M.; Halasyamani,~P.~S.; Casanova,~D.; Llunell,~M.; Alemany,~P.;
  Alvarez,~S. Distortions in {{Octahedrally Coordinated}}
  D{\textsuperscript{0}} {{Transition Metal Oxides}}:\, {{A Continuous Symmetry
  Measures Approach}}. \emph{Chemistry of Materials} \textbf{2006}, \emph{18},
  3176--3183\relax
\mciteBstWouldAddEndPuncttrue
\mciteSetBstMidEndSepPunct{\mcitedefaultmidpunct}
{\mcitedefaultendpunct}{\mcitedefaultseppunct}\relax
\EndOfBibitem
\bibitem[Catti and Ghaani(2013)Catti, and Ghaani]{catti2013}
Catti,~M.; Ghaani,~M.~R. On the Lithiation Reaction of Niobium Oxide:
  Structural and Electronic Properties of
  {{Li}}{\textsubscript{1.714}}{{Nb}}{\textsubscript{2}}{{O}}{\textsubscript{5}}.
  \emph{Physical Chemistry Chemical Physics} \textbf{2013}, \emph{16},
  1385--1392\relax
\mciteBstWouldAddEndPuncttrue
\mciteSetBstMidEndSepPunct{\mcitedefaultmidpunct}
{\mcitedefaultendpunct}{\mcitedefaultseppunct}\relax
\EndOfBibitem
\bibitem[Catti \latin{et~al.}(2015)Catti, Pinus, and Knight]{catti2015}
Catti,~M.; Pinus,~I.; Knight,~K. Lithium Insertion Properties of
  {{Li}}{\textsubscript{x}}{{TiNb}}{\textsubscript{2}}{{O}}{\textsubscript{7}}
  Investigated by Neutron Diffraction and First-Principles Modelling.
  \emph{Journal of Solid State Chemistry} \textbf{2015}, \emph{229},
  19--25\relax
\mciteBstWouldAddEndPuncttrue
\mciteSetBstMidEndSepPunct{\mcitedefaultmidpunct}
{\mcitedefaultendpunct}{\mcitedefaultseppunct}\relax
\EndOfBibitem
\bibitem[Urban \latin{et~al.}(2016)Urban, Seo, and Ceder]{urban2016}
Urban,~A.; Seo,~D.-H.; Ceder,~G. Computational Understanding of {{Li}}-Ion
  Batteries. \emph{npj Computational Materials} \textbf{2016}, \emph{2},
  npjcompumats20162\relax
\mciteBstWouldAddEndPuncttrue
\mciteSetBstMidEndSepPunct{\mcitedefaultmidpunct}
{\mcitedefaultendpunct}{\mcitedefaultseppunct}\relax
\EndOfBibitem
\bibitem[Dalton \latin{et~al.}(2012)Dalton, Belak, and {Van der
  Ven}]{dalton2012}
Dalton,~A.~S.; Belak,~A.~A.; {Van der Ven},~A. Thermodynamics of {{Lithium}} in
  {{TiO}}{\textsubscript{2}}({{B}}) from {{First Principles}}. \emph{Chemistry
  of Materials} \textbf{2012}, \emph{24}, 1568--1574\relax
\mciteBstWouldAddEndPuncttrue
\mciteSetBstMidEndSepPunct{\mcitedefaultmidpunct}
{\mcitedefaultendpunct}{\mcitedefaultseppunct}\relax
\EndOfBibitem
\bibitem[R{\"u}scher and Nygren(1991)R{\"u}scher, and Nygren]{ruscher1991}
R{\"u}scher,~C.~H.; Nygren,~M. Magnetic Properties of Phases Possessing Block
  Type Structures in the {{Nb}}{\textsubscript{2}}{{O}}{\textsubscript{5-2x}}
  System, with 0{$\leq$}x{$\leq$}0.083. \emph{Journal of Physics: Condensed
  Matter} \textbf{1991}, \emph{3}, 3997\relax
\mciteBstWouldAddEndPuncttrue
\mciteSetBstMidEndSepPunct{\mcitedefaultmidpunct}
{\mcitedefaultendpunct}{\mcitedefaultseppunct}\relax
\EndOfBibitem
\bibitem[Lee and Pickett(2015)Lee, and Pickett]{lee2015}
Lee,~K.-W.; Pickett,~W.~E. Organometalliclike Localization of
  4{\emph{d}}-Derived Spins in an Inorganic Conducting Niobium Suboxide.
  \emph{Physical Review B} \textbf{2015}, \emph{91}, 195152\relax
\mciteBstWouldAddEndPuncttrue
\mciteSetBstMidEndSepPunct{\mcitedefaultmidpunct}
{\mcitedefaultendpunct}{\mcitedefaultseppunct}\relax
\EndOfBibitem
\bibitem[Richter \latin{et~al.}(2005)Richter, Henningsson, Karlsson, Andersson,
  Uvdal, Siegbahn, and Sandell]{richter2005}
Richter,~J.~H.; Henningsson,~A.; Karlsson,~P.~G.; Andersson,~M.~P.; Uvdal,~P.;
  Siegbahn,~H.; Sandell,~A. Electronic Structure of Lithium-Doped Anatase
  {{TiO}}{\textsubscript{2}} Prepared in Ultrahigh Vacuum. \emph{Physical
  Review B} \textbf{2005}, \emph{71}, 235418\relax
\mciteBstWouldAddEndPuncttrue
\mciteSetBstMidEndSepPunct{\mcitedefaultmidpunct}
{\mcitedefaultendpunct}{\mcitedefaultseppunct}\relax
\EndOfBibitem
\bibitem[Morgan and Watson(2010)Morgan, and Watson]{morgan2010}
Morgan,~B.~J.; Watson,~G.~W. {{GGA}}+{{U}} Description of Lithium Intercalation
  into Anatase {{TiO}}{\textsubscript{2}}. \emph{Physical Review B}
  \textbf{2010}, \emph{82}, 144119\relax
\mciteBstWouldAddEndPuncttrue
\mciteSetBstMidEndSepPunct{\mcitedefaultmidpunct}
{\mcitedefaultendpunct}{\mcitedefaultseppunct}\relax
\EndOfBibitem
\bibitem[Maxisch \latin{et~al.}(2006)Maxisch, Zhou, and Ceder]{maxisch2006}
Maxisch,~T.; Zhou,~F.; Ceder,~G. Ab Initio Study of the Migration of Small
  Polarons in Olivine {{Li}}{\textsubscript{x}}{{FePO}}{\textsubscript{4}} and
  Their Association with Lithium Ions and Vacancies. \emph{Physical Review B}
  \textbf{2006}, \emph{73}, 104301\relax
\mciteBstWouldAddEndPuncttrue
\mciteSetBstMidEndSepPunct{\mcitedefaultmidpunct}
{\mcitedefaultendpunct}{\mcitedefaultseppunct}\relax
\EndOfBibitem
\bibitem[R{\"u}scher(1992)]{ruscher1992}
R{\"u}scher,~C. The Structural Effect on the Electrical Properties of
  {{NbO}}{\textsubscript{2.5-x}} Block-Type Compounds. \emph{Physica C:
  Superconductivity} \textbf{1992}, \emph{200}, 129--139\relax
\mciteBstWouldAddEndPuncttrue
\mciteSetBstMidEndSepPunct{\mcitedefaultmidpunct}
{\mcitedefaultendpunct}{\mcitedefaultseppunct}\relax
\EndOfBibitem
\bibitem[Zhong \latin{et~al.}(1992)Zhong, Dahn, and Colbow]{zhong1992}
Zhong,~Q.; Dahn,~J.~R.; Colbow,~K. Lithium Intercalation into
  {{WO}}{\textsubscript{3}} and the Phase Diagram of
  {{Li}}{\textsubscript{x}}{{WO}}{\textsubscript{3}}. \emph{Physical Review B}
  \textbf{1992}, \emph{46}, 2554--2560\relax
\mciteBstWouldAddEndPuncttrue
\mciteSetBstMidEndSepPunct{\mcitedefaultmidpunct}
{\mcitedefaultendpunct}{\mcitedefaultseppunct}\relax
\EndOfBibitem
\bibitem[Walkingshaw \latin{et~al.}(2004)Walkingshaw, Spaldin, and
  Artacho]{walkingshaw2004}
Walkingshaw,~A.~D.; Spaldin,~N.~A.; Artacho,~E. Density-Functional Study of
  Charge Doping in {{WO}}{\textsubscript{3}}. \emph{Physical Review B}
  \textbf{2004}, \emph{70}, 165110\relax
\mciteBstWouldAddEndPuncttrue
\mciteSetBstMidEndSepPunct{\mcitedefaultmidpunct}
{\mcitedefaultendpunct}{\mcitedefaultseppunct}\relax
\EndOfBibitem
\bibitem[Ohno \latin{et~al.}(2019)Ohno, Helm, Fuchs, Dewald, Kraft, Culver,
  Senyshyn, and Zeier]{ohno2019}
Ohno,~S.; Helm,~B.; Fuchs,~T.; Dewald,~G.; Kraft,~M.~A.; Culver,~S.~P.;
  Senyshyn,~A.; Zeier,~W.~G. Further {{Evidence}} for {{Energy Landscape
  Flattening}} in the {{Superionic Argyrodites
  Li}}{\textsubscript{6+{\emph{x}}}}{{P}}{\textsubscript{1\textendash{\emph{x}}}}{{M}}{\textsubscript{{\emph{x}}}}{{S}}{\textsubscript{5}}{{I}}
  ({{M}} = {{Si}}, {{Ge}}, {{Sn}}). \emph{Chemistry of Materials}
  \textbf{2019}, \emph{31}, 4936--4944\relax
\mciteBstWouldAddEndPuncttrue
\mciteSetBstMidEndSepPunct{\mcitedefaultmidpunct}
{\mcitedefaultendpunct}{\mcitedefaultseppunct}\relax
\EndOfBibitem
\end{mcitethebibliography}

\end{document}